\documentclass[useAMS,usenatbib]{mn2e}
\usepackage{Times}
\usepackage{psfrag}
\usepackage{pspicture}
\usepackage{graphicx}
\usepackage{amsmath}

\def\gs{\mathrel{\raise0.35ex\hbox{$\scriptstyle >$}\kern-0.6em\lower0.40ex\hbox{{$\scriptstyle \sim$}}}} 
\def\ls{\mathrel{\raise0.35ex\hbox{$\scriptstyle <$}\kern-0.6em\lower0.40ex\hbox{{$\scriptstyle \sim$}}}}

\def\Wm2{\,\hbox{W}\,\hbox{m}^{-2}} 
\def\gsim{\mathrel{\raise0.35ex\hbox{$\scriptstyle >$}\kern-0.6em\lower0.40ex\hbox{{$\scriptstyle \sim$}}}} 
\def\lsim{\mathrel{\raise0.35ex\hbox{$\scriptstyle <$}\kern-0.6em\lower0.40ex\hbox{{$\scriptstyle \sim$}}}} 
\def\ltsima{$\; \buildrel < \over \sim \;$} 
\def\simlt{\lower.5ex\hbox{\ltsima}} 
\def\gtsima{$\; \buildrel > \over \sim \;$} 
\def\simgt{\lower.5ex\hbox{\gtsima}}

\title[Boosted SF \& AGN activity in a massive merging cluster]{MC$^2$: Boosted AGN and star-formation activity in CIZA J2242.8+5301, a massive post-merger cluster at $\bf z=0.19$\thanks{Based on observations obtained with WFC on the INT, programmes I12BN003, I13BN006 and I13BN008; AF2+WYFFOS on the WHT, programme W14AN012 and on DEIMOS/Keck observations under program U156D.}}
\author[D. Sobral et al.]{David Sobral$^{1,2,3}$\thanks{FCT-IF/Veni Fellow. E-mail: sobral@iastro.pt}, Andra Stroe$^{3}$, William A.\, Dawson$^{4,5}$, David Wittman$^{5}$, James Jee$^{5}$,\newauthor  Huub R{\"o}ttgering$^{3}$, Reinout J. van Weeren$^{6}$, Marcus Br{\"u}ggen$^{7}$ \\
$^{1}$ Instituto de Astrof\'{\i}sica e Ci\^{e}ncias do Espa\c{c}o, Universidade de Lisboa, OAL, Tapada da Ajuda, PT1349-018 Lisbon, Portugal \\
$^{2}$ Departamento de F\'{i}sica, Faculdade de Ci\^{e}ncias, Universidade de Lisboa, Edif\'{i}cio C8, Campo Grande, PT1749-016 Lisbon, Portugal \\
$^{3}$ Leiden Observatory, Leiden University, P.O.\ Box 9513, NL-2300 RA Leiden, The Netherlands \\
$^{4}$ Lawrence Livermore National Laboratory, 7000 East Ave, Livermore, CA 94550, USA\\
$^{5}$ Department of Physics, University of California, Davis, One Shields Avenue, Davis, CA 95616, USA\\
$^{6}$ Harvard-Smithsonian Center for Astrophysics, 60 Garden Street, Cambridge, MA 02138, USA \\
$^{7}$ Hamburger Sternwarte, Gojenbergsweg 112, 21029 Hamburg, Germany
}

\begin{document}
\date{LLNL-JRNL-661314-DRAFT. Accepted 2015 March 6. Received 2015 March 6; in original form 2015 Jan19}
\pagerange{\pageref{firstpage}--\pageref{lastpage}} \pubyear{2015}

\maketitle
\label{firstpage}
\begin{abstract}
\noindent Cluster mergers may play a fundamental role in the formation and evolution of cluster galaxies. \cite{Stroe14a} revealed unexpected over-densities of candidate H$\alpha$ emitters near the $\sim1$\,Mpc-wide shock fronts of the massive ($\sim2\times10^{15}$\,M$_{\odot}$) ``Sausage'' merging cluster, CIZA J2242.8+5301. We used Keck/DEIMOS and WHT/AF2 to confirm 83 H$\alpha$ emitters in and around the merging cluster. We find that cluster star-forming galaxies in the hottest X-ray gas and/or in the cluster sub-cores (away from the shock fronts) show high [S{\sc ii}]\,6716/[S{\sc ii}]\,6761 and high [S{\sc ii}]\,6716/H$\alpha$, implying very low electron densities ($<30\times$ lower than all other star-forming galaxies outside the cluster) and significant contribution from supernovae, respectively. All cluster star-forming galaxies near the cluster centre show evidence of significant outflows (blueshifted Na\,D\,$\sim200-300$\,km\,s$^{-1}$), likely driven by supernovae. Strong outflows are also found for the cluster H$\alpha$ AGN. H$\alpha$ star-forming galaxies in the merging cluster follow the $z\sim0$ mass-metallicity relation, showing systematically higher metallicity ($\sim$0.15-0.2\,dex) than H$\alpha$ emitters outside the cluster (projected $R>2.5$\,Mpc). This suggests that the shock front may have triggered remaining metal-rich gas which galaxies were able to retain into forming stars. Our observations show that the merger of impressively massive ($\sim10^{15}$\,M$_\odot$) clusters can provide the conditions for significant star-formation and AGN activity, but, as we witness strong feedback by star-forming galaxies and AGN (and given how massive the merging cluster is), such sources will likely quench in a few 100\,Myrs.

\end{abstract}

\section{Introduction}

Star-forming galaxies have evolved dramatically in the 11\,Gyr between $z\sim2.5$ (the likely peak of the star formation history of the Universe) and the present day \citep[e.g.\ ][]{Madau96,Sobral09,Karim11,Sobral14}. The co-moving star formation rate density of the Universe has dropped by more than an order of magnitude over this time in all environments \citep{Rodighiero11,Karim11,Gilbank11,Sobral13}, and also specifically in clusters \citep[e.g.][]{Kodama13,Shimakawa14}. The bulk of this evolution is described by the continuous decrease of the typical star formation rate, SFR$^*$, which is found to affect the star-forming population at all masses \citep{Sobral14}. Surprisingly, the decline of SFR$^*$ seems to be happening (for star-forming galaxies) in all environments, at least since $z\sim2$ \citep[e.g.][]{Koyama13}.

Locally, star formation activity has been found to be very strongly dependent on environment \citep[e.g.][]{lewiss,Gomez,Tanaka04,Mahajan2010}. Clusters of galaxies are dominated by passively-evolving galaxies, while star-forming galaxies are mostly found in low-density/field environments \citep{Dressler80}. It is also well-established \citep[e.g.][]{Gomez,Kauff,Best2004} that both the typical star formation rates of galaxies and the star-forming fraction decrease with local environmental density both in the local Universe and at moderate redshift \citep[$z\sim0.4$, e.g.][]{Kodama04}. This is in line with the results at $z\sim0.2-0.3$ from \cite{Couch01} or \cite{Balogh02} who found that the H$\alpha$ luminosity (an excellent tracer of recent star-formation activity) function in rich, relaxed clusters have the same shape as in the field, but have a much lower normalisation ($\sim50$ per cent lower), consistent with a significant suppression of star formation in highly dense environments. 

The strong positive correlation between star formation rate (SFR) and stellar mass \citep[e.g.][]{Brinchmann04,Noeske2007,Peng}, while being a strong function of cosmic time/redshift, seems to depend little on environment \citep{Koyama13}, even though cluster star-forming galaxies seem to be more massive than field star-forming galaxies. Thus, the fundamental difference between cluster and field environments (regarding their relation with star formation) seems to be primarily the fraction of star-forming galaxies, or the probability of being a star-forming galaxy: it is much lower in cluster environments than in field environments. Studies looking at the mass-metallicity relation with environment also seem to find relatively little difference at $z\sim1$ \citep[comparing groups and fields;][]{Sobral13B}, or just a slight offset ($+0.04$ dex) for relaxed cluster galaxies in the Local Universe, as compared to the field \citep[using Sloan Digital Sky Survey, SDSS;][]{Ellison09}. Further studying the mass-metallicity relation \citep[and the Fundamental Metallicity Relation, FMR, e.g.][]{Maiolino08,Mannucci10,Stott13b} in clusters and comparing to the field could provide further important information.
 
While there are increasing efforts to try to explain the SFR dependence on the environment, by conducting surveys at high redshift \citep[e.g.][]{Hayashi,Sobral11,Matsuda11,Muzzin12,Koyama13,Tal2014,Darvish14}, so far such studies have not been able to fully reveal the physical processes leading to the ultimate quenching of (satellite) star-forming galaxies \citep[e.g.][]{Peng,Muzzin12,Muzzin2014}. Several strong processes have been proposed and observed, such as harassment \citep[e.g.][]{Moore1996}, strangulation \citep[e.g.][]{Larson1980} and ram-pressure stripping \citep[e.g.][]{Gunn1972,Fumagalli2014}. Observations are also showing a variety of blue-shifted rest-frame UV absorption lines which indicate that most star-forming galaxies at least at $z\sim1-2$, are able to drive powerful gas outflows \citep[e.g.][]{Shapley03,Weiner09,Kornei12} which may play a significant role in quenching, particularly if those happen in high density environments. Evidence of such galactic winds have also been seen in e.g. \cite{FSchreiber09} through broad components in the rest-frame optical H$\alpha$ and [N{\sc ii}] emission line profiles \citep[e.g.][]{Genzel11}. Spatially resolved observations allow for constraints on the origin of the winds within galaxies, and on the spatial extent of the outflowing gas, which are essential to derive mass outflow rates. In field environments, it is expected that such outflows will not be able to escape the halo (as long as it is massive enough and it is not a satellite), and in many conditions would likely come back and further fuel star formation \citep[e.g.][]{Hopkins13}. However, in the most massive clusters, such strong outflows will likely result in significant amounts of gas being driven out of the sub-halos that host star-forming galaxies, enriching the ICM and quickly quenching star-forming galaxies (SFGs) with the highest SFRs/highest outflow rates.

Many studies often have environmental classes simply divided into (relaxed) ``clusters" or ``fields". However, in a $\Lambda$CDM Universe, most clusters are expected to be the result of group/smaller cluster mergers -- some of which can be extremely violent. Little is known about the role of cluster and group mergers in galaxy formation and evolution, and whether they could be important in setting the environmental trends which have now been robustly measured and described. It is particularly important to understand if cluster mergers trigger star formation \citep[e.g.][]{Miller2003,Owen2005,Ferrari2005,Hwang2009,Wegner2015}, if they quench it \citep[e.g.][]{Poggianti04}, or, alternatively, if they have no direct effect \citep[e.g.][]{Chung10}. Results from \cite{Umeda04}, studying a merging cluster at $z\sim0.2$ (Abell 521) found tentative evidence that merging clusters could perhaps trigger star-formation. More recently, \cite{Stroe14a} conducted a wide field H$\alpha$ narrow-band survey over two merging clusters with a simple geometry, with the merger happening in the plane of the sky. \cite{Stroe14a} find a strong boost in the normalisation of the H$\alpha$ luminosity function of the CIZA J2242.8+5301 (``Sausage") cluster, several times above the field and other clusters. The authors suggest that they may be witnessing star-formation enhancement or triggered due to the passage of the shock wave seen in the radio and X-rays. Interestingly, \cite{Stroe14a} do not find this effect on the other similar merging cluster studied (``Toothbrush"), likely because it is a significantly older merger \citep[about 1Gyr older, c.f.][]{Stroe14a,Stroe15}, and thus displays only the final result (an excess of post-starburst galaxies instead of H$\alpha$ emitters). The results are in very good agreement with simulations by \cite{Roediger14} and recent observational results by \cite{Pranger14}.

In order to investigate the nature of the numerous H$\alpha$ emitter candidates in and around the ``Sausage" merging cluster, we have obtained deep spectroscopic observations of the bulk of the sample presented in \cite{Stroe15}, using Keck/DEIMOS \citep[PI Wittman;][]{Dawson2014} and the William Herschel Telescope (WHT) AutoFib2+WYFFOS (AF2) instrument (PI: Stroe; this paper). In this paper, we use these data to confirm candidate H$\alpha$ emitters, unveil their nature, masses, metallicities and other properties. We use a cosmology with $\Omega_{\Lambda}$\,=\,0.7, $\Omega_{m}$\,=\,0.3, and H$_{0}$\,=\,70\,km\,s$^{-1}$\,Mpc$^{-1}$. All quoted magnitudes are on the AB system and we use a Chabrier initial mass function \citep[IMF;][]{Chabrier}.

\section{Sample Selection, Observations \& Data Reduction}

\subsection{The ``Sausage" Merging Cluster}

\begin{figure*}
  \centering
  \includegraphics[width=18cm]{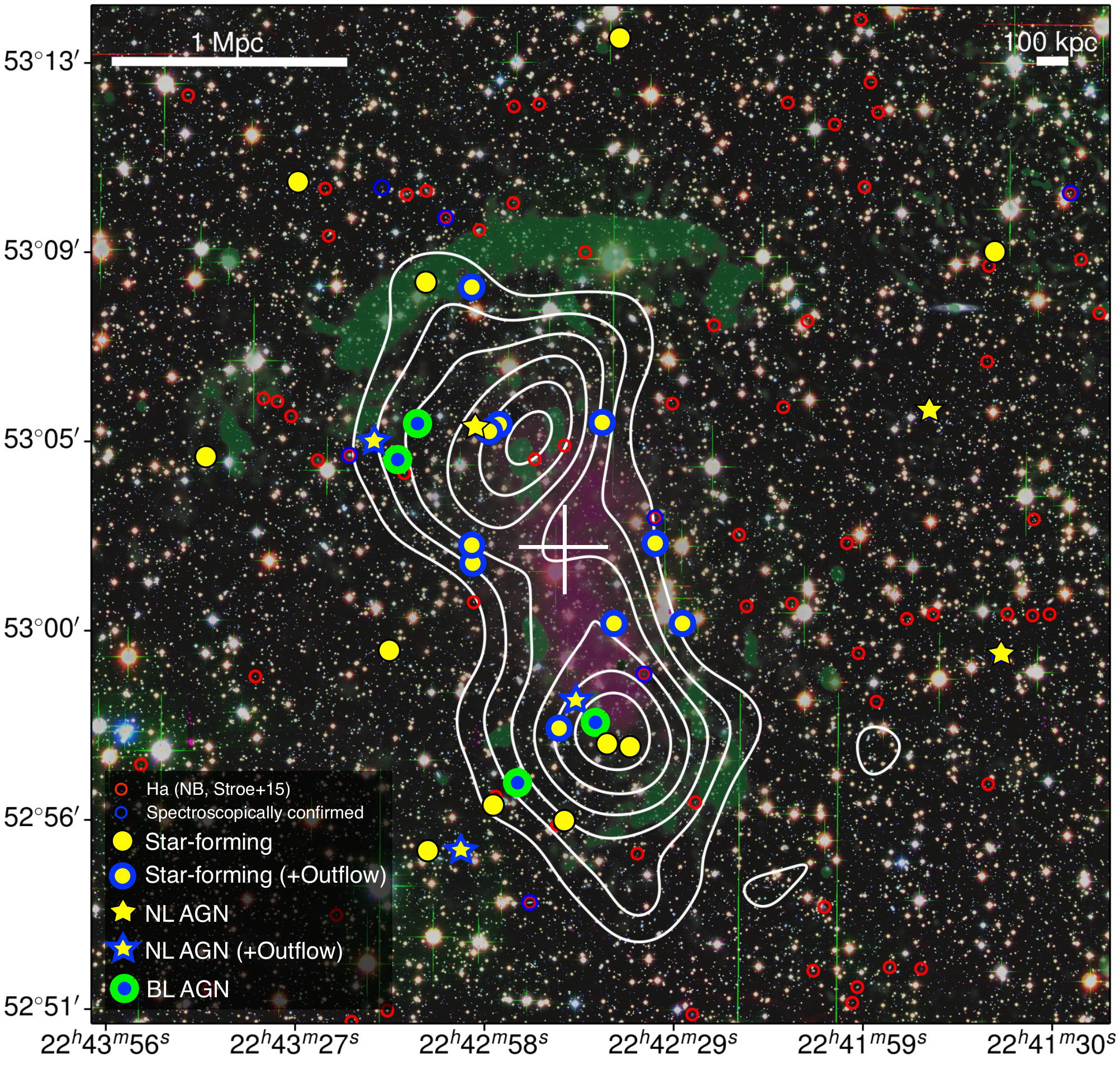}
\caption{The distribution of H$\alpha$ emitters in the Sausage cluster. The cross marks what we define as the ``centre" of the cluster. The background shows a false {\sc rgb} image from the combination of broad-band images presented in \citealt{Stroe15}, while white contours present the weak lensing map \citep{Jee2015} and in green the 323\,MHz radio emission \citep[][]{Stroe13}. The H$\alpha$ emitters in our sample reside in a range of different regions, but are found preferably near the shock fronts. H$\alpha$ emitters also seem to be found just on the outskirts of the hottest X-ray gas -- but where the temperatures are still very high \citep[][]{Ogrean13}. We also find that all AGN are located relatively close to the post-shock front, but all at a couple of hundred projected kpc away from the radio relics.  Note that our sample extends beyond this region, as the field of view of the narrow-band survey, and the spectroscopic follow-up of such sources, cover a larger area \citep[see][]{Stroe15}. We also show star-forming galaxies showing signatures of outflows, mostly from systematically blue-shifted Na\,D in absorption from 150 to 300\,km$^{-1}$. Note that 100\% of the cluster star-forming galaxies which are closest to the hottest X-ray gas (very close to the ``centre" of the cluster), have strong signatures of outflows. Potentially, these are also the sources that, if affected by the shock, may have been the first to be affected, up to $\sim0.7$\,Gyr ago.}
\label{SAUSAGE_SKY}
\end{figure*}

The CIZA J2242.8+5301 cluster (nicknamed  ``Sausage'' cluster, referred simply as Sausage for the rest of the paper; see Figure  \ref{SAUSAGE_SKY}) is a $z=0.1921$, X-ray luminous \citep[$L_{0.1-2.4\mathrm{keV}}=6.8\times10^{44}$ erg s$^{-1}$;][]{Kocevski07}, disturbed \citep[][Akamatsu et al. 2015]{Akamatsu13,Ogrean13,Ogrean14} cluster that hosts double radio relics towards its northern and southern outskirts \citep[][see Figure \ref{SAUSAGE_SKY}]{vanWeeren10,Stroe13}. The radio relics (Mpc-wide patches of diffuse radio emission) trace Mpc-wide shock fronts travelling through the intra-cluster medium (see Figure \ref{SAUSAGE_SKY}) thought to have been produced at the core-passage of two massive clusters during major merger in the plane of the sky \citep[][]{vanWeeren10,Stroe13,Stroe14c,Stroe14b}. Despite being an interesting cluster, the CIZA J2242.8+5301 cluster (Sausage cluster from now on) remained mostly unexplored until very recently, due to significant Galactic extinction \citep[c.f.][]{Stroe14a,Jee2015,Stroe15}.

Dynamics inferred from spectroscopic observations suggest the two sub-clusters each have masses of $\sim1.3-1.6\times 10^{15} M_\odot$ \citep[][]{Dawson2014}, in agreement with independent weak lensing analysis which points towards $\sim 1.0-1.1\times 10^{15} M_\odot$ \citep{Jee2015}. The weak lensing \citep{Jee2015}, and the dynamics \citep[][]{Dawson2014} point towards a total mass of $\approx 2\times 10^{15}$\,M$_\odot$, making it one of the most massive clusters known to date. The virial radius for the total system from weak lensing \citep{Jee2015} is r$_{200}\sim2.63$\,Mpc..

\cite{Dawson2014} presents a detailed dynamics analysis of the cluster merger. Observations and information from lensing, spectroscopy, broad-band imaging, radio and other constraints imply that the merger likely happened around 0.7$\pm0.2$\,Gyrs ago (see also Akamatsu et al. 2015, in very good agreement). Clusters were likely travelling at a velocity of $\sim2000-2200$\,km\,s$^{-1}$ towards each other when they merged \citep[][Akamatsu et al. 2015]{Dawson2014}. This is in excellent agreement with the analysis presented in \cite{Stroe14c} that shows that the shock wave seems to be moving with a similar speed ($\sim2000-2500$\,km\,s$^{-1}$). Because the shock does not slow down due to gravitational effects, it can be thought as a proxy of the collisional velocity, further supporting a speed of $\sim2000$\,km\,s$^{-1}$ (see also Akamatsu et al. 2015 who find this is also the case from X-rays). We use the detailed information from \cite{Dawson2014}, \cite{Jee2015}, Akamatsu et al. (2015), \cite{Stroe15}, and references therein, to put our results into context and to explore potential interpretations of the results. The reader is referred to those papers for more information on the cluster itself.

\subsection{Narrow-band survey and the sample of H$\alpha$ candidates} \label{NBsurvey}

By using a custom-designed narrow-band filter ($\lambda=7839\pm55${\AA}, PI: Sobral) mounted on the Wide Field Camera at the prime-focus of the Isaac Newton Telescope, \cite{Stroe14a} imaged the Sausage cluster over $0.3$ deg$^2$ and selected $181$ potential line emitters, down to a H$\alpha$ luminosity of $10^{40.8}$ erg s$^{-1}$ \citep[see][]{Stroe14a}. They discover luminous, extended, tens-of-kpc-wide candidate H$\alpha$ emitters in the vicinity of the shock fronts, corresponding to a significant boost in the normalisation of the H$\alpha$ luminosity function, when comparing to not only the field environment \citep{Shioya,Drake13}, but also to other relaxed and merging clusters \citep[e.g.][]{Umeda04}.

%
%
%
%

\cite{Stroe15} presents deeper narrow-band and $i$ band imaging, along with new multi-band data ($BVriZ$), and find a total of 201 candidate line emitters. Here we use the full sample of candidate line emitters in and around the Sausage merging cluster, without any pre-selection on the likelihood of them being H$\alpha$, along with the corrected broad-band photometry \citep[due to Galactic dust extinction, see][]{Stroe15}. We take this approach in order to increase the completeness of our H$\alpha$ sample and avoid any biases (even if small) caused by the need to use broad-band colours and/or photometric redshifts (photo-$z$s). Spectroscopic redshifts obtained here are used in \cite{Stroe15} to test their selection, improve completeness, and reduce contamination by other emission lines.

\subsection{Follow-up spectroscopy with Keck and WHT} \label{spec_observations}

\subsubsection{Keck/DEIMOS observations} \label{DEIMOS_observations}


We conducted a spectroscopic survey of the Sausage cluster with the DEep Imaging Multi-Object Spectrograph \citep[DEIMOS;][]{Faber:2003ev} on the Keck II 10\,m telescope over two observing runs on 2013 July 14 and 2013 September 5. For full details on the observations and data reduction, see \cite{Dawson2014}. Here we provide just a brief summary.

We observed a total of four slit masks with approximately 120 slits per mask. For each mask we took three 900\,s exposures, for a total exposure time of 2.7\,ks. The average seeing was approximately 0.7$''$. For both observing runs we used 1\arcsec \ wide slits with the 1200\,line\,mm$^{-1}$ grating, tilted to a central wavelength of 6700\,\AA, resulting in a pixel scale of 0.33\,\AA\,pixel$^{-1}$, a resolution of $\sim1$\,\AA \ ($\sim45$\,km\,s$^{-1}$, observed and just below 40\,km\,s$^{-1}$ rest-frame for our cluster H$\alpha$ emitters), and typical wavelength coverage of 5400\,\AA \ to 8000\,\AA. The actual wavelength coverage is in practice shifted by $\sim\pm400$\,\AA \ depending where the slit is located along the width of the slit-mask. For most cluster members this enabled us to observe H$\beta$, [O{\sc iii}]$_{4959 \& 5007}$, MgI (b), FeI,  NaI (D) , [O{\sc i}], H$\alpha$, [N{\sc ii}] and [S{\sc ii}]. We used the DEEP2 version of the {\sc spec2d} package \citep{Newman:2012ta} to reduce the data. {\sc spec2d} performs wavelength calibration, cosmic ray removal and sky subtraction on slit-by-slit basis, generating a processed two-dimensional spectrum for each slit.  The {\sc spec2d} pipeline also generates a processed one-dimensional spectrum for each slit. This extraction creates a one-dimensional spectrum of the target, containing the summed flux at each wavelength in an optimised window.

\begin{figure*}
  \centering
 \includegraphics[width=16.2cm]{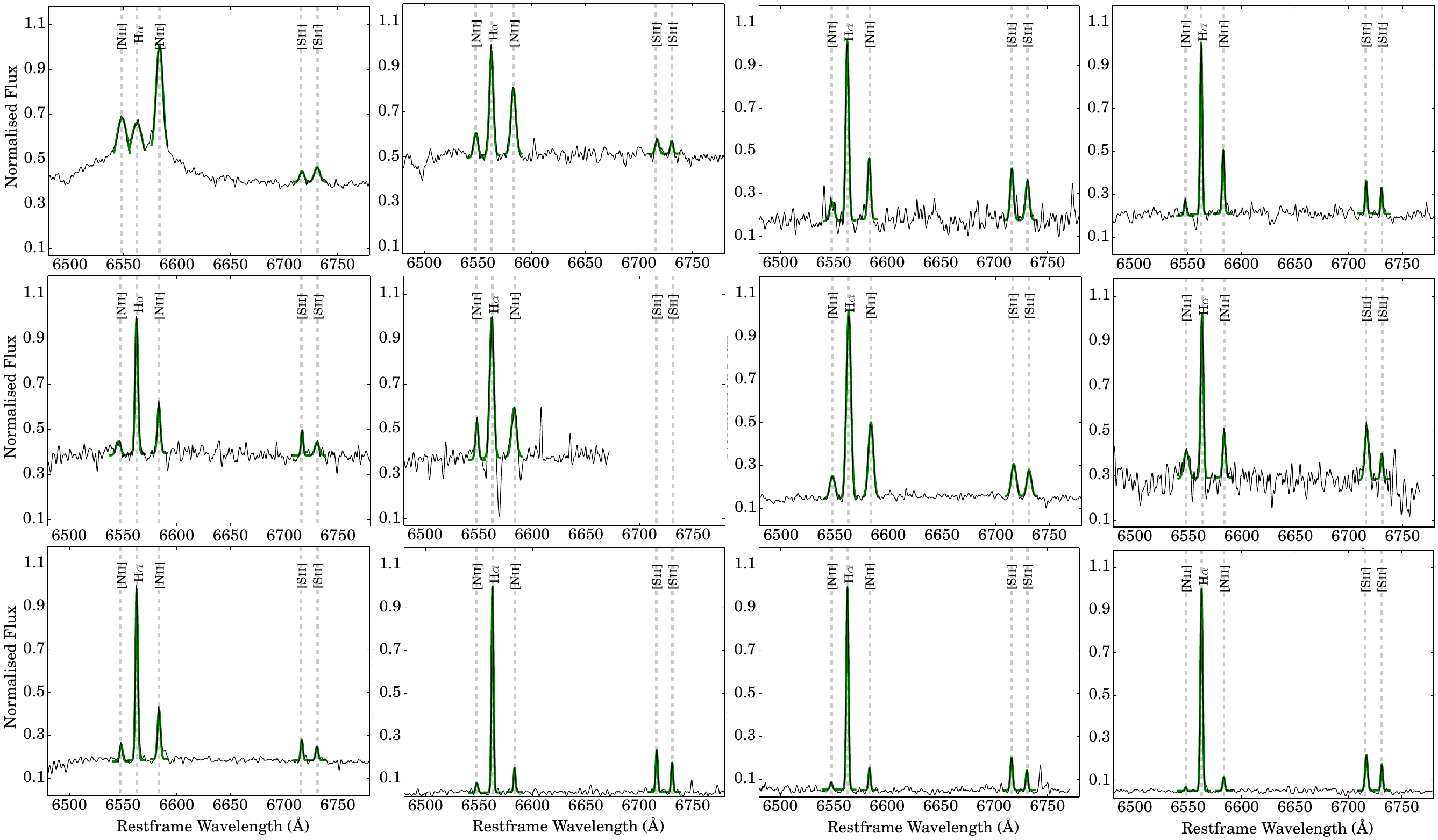}
\caption{Some examples of our H$\alpha$ emitters from the Keck/DEIMOS data and the Gaussian fits we derived in order to measure emission line ratios. We find a variety of line ratios and FWHMs, but are able to fit all emission lines with simple Gaussian profiles.}
\label{SPECTRA_HA}
\end{figure*}

Our primary DEIMOS targets were candidate red sequence/cluster galaxies and for details on the full sample, the reader is referred to \cite{Dawson2014}. Here we focus on the observed 40 H$\alpha$ emitters within the DEIMOS data-set (see e.g. Figure \ref{SPECTRA_HA}), out of which 32 are found to be cluster members \citep[see][]{Dawson2014}. The remaining 8 sources were found to be at slightly higher and slightly lower redshifts, and will be used as part of the comparison sample (H$\alpha$ emitters outside the cluster, which are either at a different redshift from the cluster, or are at a projected distance higher than 2.5\,Mpc from the cluster).

%
%
%
%

\subsubsection{WHT/AF2 observations} \label{observation_AF2}


We followed up 103 candidate line emitters from \cite{Stroe15} using AF2 on the WHT in La Palma on two nights during 2014 July 2--3. In order to allocate spare fibres, we used our $BVriZ$ photometric catalogue \citep{Stroe15} to select other potential cluster candidates (using colour-colour selections; see Stroe et al. 2015). We observed six of these sources. We found no evidence of emission lines in any of these sources, but have a very high success rate in detecting emission lines for the main sample of emission line candidates.

The seeing was 0.8-1.0$''$ throughout the observing run. The AF2 instrument on WHT is made of $\sim150$ fibres, each with a diameter of 1.6$''$, which can be allocated to sources within a $\sim30\times30$\,arcmin$^2$ field of view, although with strong spatial constraints/limitations. The spectral coverage varies slightly depending on the fibre and field location, but for a source at $z=0.19$ all our spectra cover the main emission lines we are interested in: H$\beta$, [O{\sc iii}], H$\alpha$, [N{\sc ii}] and [S{\sc ii}]. We obtained 2 different pointings: one centred on the cluster, with a total exposure time of 9\,ks (where we were able to allocate 63 fibres to targets, and 3 fibres to sky), and one slightly to the North, with a total exposure time of 5.4\,ks (46 fibres allocated to targets and 4 to sky). We also obtained some further sky exposures to improve the sky subtraction (2.7\,ks per field).

We took standard steps in the reduction of optical multi-fibre spectra, also mimicking the steps followed for DEIMOS. Biases and lamp flats were taken at the beginning of each night. Arcs using neon, helium and mercury lamps were taken on the sky for each fibre configuration. The traces of the fibres on the CCD were curved in the dispersion direction (y axis on the CCD). The lamp flats were used to correct for this distortion. Each fibre shape was fit with a Y pixel coordinate polynomial as function of X coordinate. All CCD pixels were corrected according to the polynomial for the closest fibre. This was done separately for each configuration, on the biases, flats, lamp arcs and the science data.

The final 2D bias subtracted and curvature corrected frames were then sky subtracted using the sky position exposure(s). In order to improve the sky subtraction we also used sky-dedicated fibres (which observe sky in all positions) to scale the counts. We further obtained the best scaling factor by minimising the residuals after sky subtraction. After subtracting the sky, we extracted sources along the dispersion axis, summing up the counts. We obtained a first order wavelength calibration by using the arcs and obtain a final wavelength calibration per fibre by using the wealth of sky lines on that particular fibre. This gives a wavelength calibration with an error (rms) of less than 1\,\AA.

%
%
%
%
\begin{figure}
 \centering
  \includegraphics[width=3.4in]{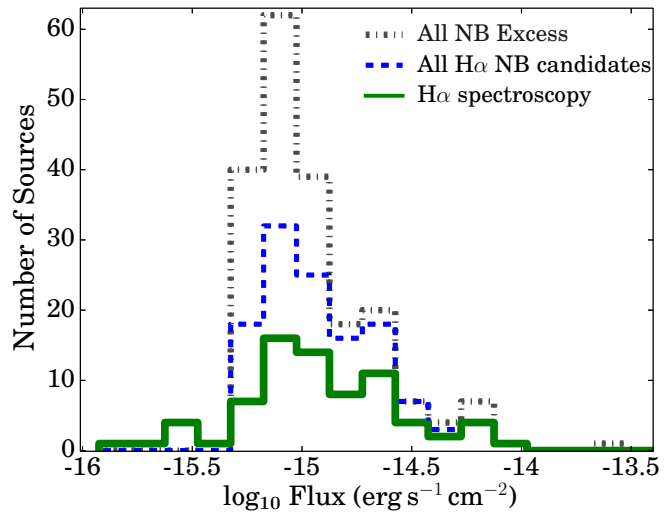}
\caption{The distribution of emission line fluxes in the full sample of narrow-band selected H$\alpha$ emitters, and those in our spectroscopic sample. Fluxes shown here are derived from narrow-band photometry (full flux, including both H$\alpha$ and [N{\sc ii}] fluxes), and corrected for Galactic extinction. This shows that we are complete down to $\sim6\times10^{-16}$\,erg\,s$^{-1}$\,cm$^{-2}$, but that, particularly due to the use of Keck, we also have H$\alpha$ emitters with significantly lower emission line fluxes in our sample, although we are clearly not complete for those fluxes.}
\label{flux_dist}
\end{figure}

In total, out of the 109 targeted sources, we obtained high enough S/N to determine a redshift for 73 sources (65 candidate line emitters selected with the NB). The remaining sources either had very low S/N, were targeted by fibres with low throughput and/or for which sky subtraction was only possible with the dedicated (different) sky fibre (thus resulting in poor sky subtraction). All the sources for which we did not get high enough S/N to detect an emission line are the emitters with the lowest emission line fluxes, expected to remain undetected with the achieved flux limit. Figure \ref{flux_dist} shows the distribution of fluxes for the full sample of candidate H$\alpha$ emitters (only a fraction of those were targeted) and those we have detected at high S/N -- this shows that we are complete for ``intrinsic" (i.e., after correcting for Galaxy extinction) fluxes of $>6\times10^{-16}$\,erg\,s$^{-1}$\,cm$^{-2}$ (see Figure \ref{flux_dist}). We note that while our the Keck spectroscopy was targeting red sequence galaxies \citep[see][]{Dawson2014}, our WHT follow-up was specifically targeted at NB-selected line emitter candidates (dominated by H$\alpha$ at $z\sim0.19$), thus giving an unbiased spectroscopic sample to study H$\alpha$ emitters. Most importantly, our AF2 sample targets line emitters both in and around the merging cluster, thus allowing us a direct comparison between cluster H$\alpha$ emitters and those outside the cluster, observed with the same instrument/configuration/exposure times.

\subsection{Redshifts and Emission line Measurements} \label{Redshifts}

%
%
%
%
%
%
%

We extract the 1D spectra (e.g. Figure \ref{SPECTRA_HA}) by detecting the high S/N trace (continuum), or by detecting the strong emission lines, and extracting them across the exposed pixels. We obtain a reasonable flux calibration with broad-band photometry available from $g$, $r$, and $i$ observations and improve it further by using our own NB observations. However, we note that the focus of this paper is on line ratios (which do not depend on flux calibration), not emission line fluxes.

\begin{figure}
 \centering
  \includegraphics[width=3.4in]{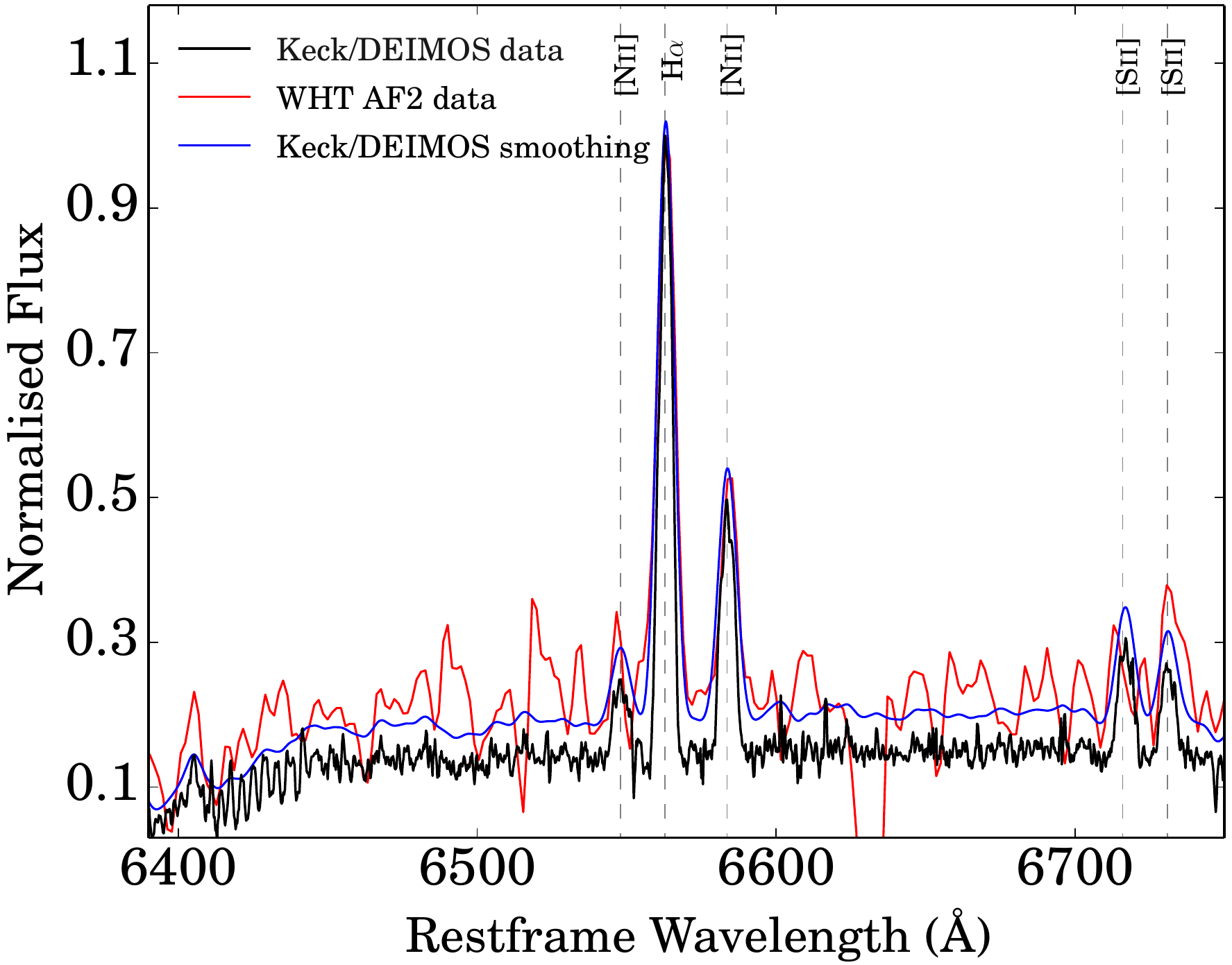}
\caption{An example of one of the typical/faint sources that was targeted with both WHT/AF2 and Keck/DEIMOS. We find perfect agreement and recover the same line ratios, within the errors, although Keck/DEIMOS spectra have much higher individual S/N ratio and much higher resolution, as this clearly shows. Nevertheless, both data-sets provide consistent measurements with no biases and thus can be used together.}
\label{AF2_DEIMOS}
\end{figure}

Spectroscopic redshifts for the Keck/DEIMOS data-set are obtained as described in \cite{Dawson2014}. We find 40 H$\alpha$ emitters within the DEIMOS/Keck data-set, but 5 (3) are at higher (lower) redshift, and thus clearly outside the merging cluster. These will be part of our comparison/field sample together with the AF2 spectra at the same redshift of the cluster but far away from it (non-cluster members). In total, 32 H$\alpha$ emitters are cluster members within the DEIMOS data-set. From these, 6 were targeted with both DEIMOS and AF2 and show perfect agreement in the redshift determination \citep[see][for a redshift comparison]{Dawson2014}, flux and emission line ratios, showing that no systematics are affecting our analysis, and that spectra from both instruments are fully comparable -- see Figure \ref{AF2_DEIMOS}.

%
%
%
%
%

%
%
%
%
%
\begin{figure}
 \centering
  \includegraphics[width=3.4in]{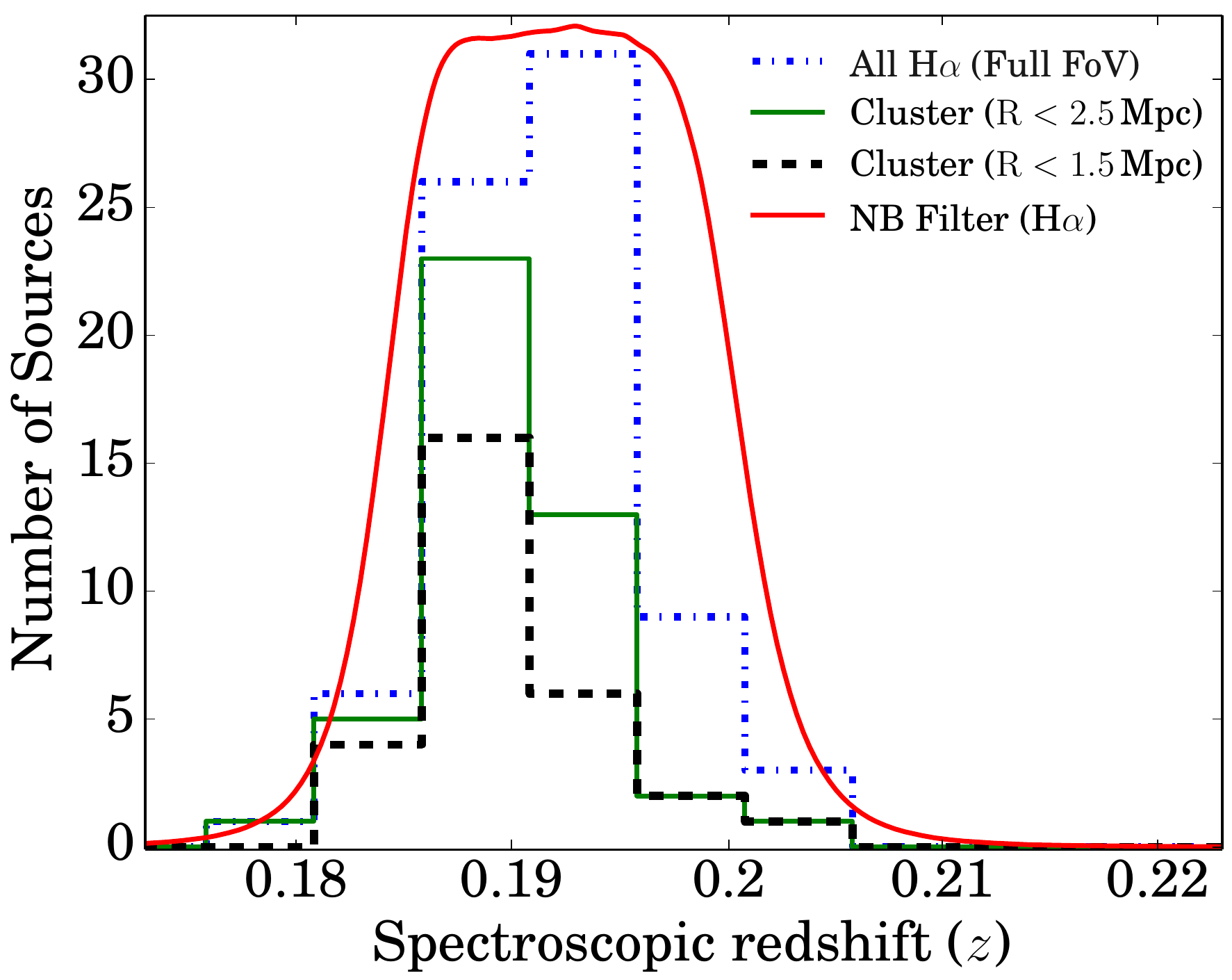}
\caption{Spectroscopic redshift distribution of our sample of H$\alpha$ emitters in the full FoV of the INT survey \citep{Stroe15}, compared with the spectroscopic distribution of H$\alpha$ emitters within the projected ``central'' 1.5\,Mpc radius (physical) of the Sausage cluster and those within a 2.5\,Mpc radius (physical) of the ``centre" of the Sausage cluster. We also scale our narrow-band filter transmission. Sources outside the 2.5\,Mpc radius are used as a comparison sample, together with a few sources at significantly higher and lower redshift found with DEIMOS (not shown here). We note that our narrow-band filter profile encompasses the 95\% confidence interval of the full cluster redshift dispersion \citep{Dawson2014} for H$\alpha$ emission.}
\label{zspec_dist}
\end{figure}

For the WHT/AF2 data-set, we determine an initial estimate for each redshift by identifying strong emission lines around $\sim7600-8000$\AA. In most cases emission lines are detected at high S/N ($>10$) and a redshift is then found with several emission lines, with the vast majority of sources being at $z\sim0.19$ with strong H$\alpha$ emission (see Figure \ref{zspec_dist} for the redshift distribution of H$\alpha$ emitters). Given the proximity to the Galaxy \citep[see e.g.][]{Jee2015}, the stellar density is many times higher than in a typical extragalactic field. We find objects with many clear absorption features which are easily classified as stars by identifying $z=0$ absorption lines (including H$\alpha$). The complete set of redshifts of H$\alpha$ emitters in our sample is given in Table \ref{table:Gal_props}.

Out of the 73 (65 line emitters) sources with high enough S/N we find 49 H$\alpha$ emitters at $z\sim0.19$, 8 H$\beta$/[OIII] emitters at $z\sim0.6$, 2 [OII] emitters at $z\sim1.1$ and one 4000\,\AA \ break galaxy at $z\sim0.8$. In total, for the AF2 spectra, we find 5 stars among our full sample of candidate line emitters. All other sources that were targeted and that were not in our NB-selected catalogue were found to be stars. Thus, within the sample of line emitters from \cite{Stroe15}, we find that 75\,\% are H$\alpha$ emitters.

Emission line fluxes for both AF2 and DEIMOS spectra are measured by fitting Gaussian profiles (see e.g. Figure \ref{SPECTRA_HA}), and measuring the continuum directly red-ward and blue-ward of the lines (masking any other features or nearby lines). We also obtain the line FWHM (in km\,s$^{-1}$), taking advantage of the high resolution, high S/N Keck spectra. We measure (observed, aperture/slit/fibre corrected) line fluxes in the range $1.7-35\times10^{-16}$\,erg\,s$^{-1}$\,cm$^{-2}$ in H$\alpha$, and FWHMs of 40-466 km\,s$^{-2}$ -- full details are given in Table \ref{table:Gal_props}. We find the best redshift by fitting all the available spectral lines and do this independently on the Keck/DEIMOS and WHT/AF2 data-sets. Given that we have an overlap of six sources, we can check if the different resolution and the use of fibres versus slits can introduce any biases/differences. We find that all these six sources yielded the same redshift and we find that the fluxes and the line ratios all agree within the errors (see an example in Figure \ref{AF2_DEIMOS}). We therefore combine the samples for the following analysis, taking into account the different errors given by each data-set. For the six sources with measurements in both data-sets we use the DEIMOS results for four out of the six sources (due to a much higher S/N). For the remaining two (detected at very high S/N in AF2), we use the AF2 measurements because they also cover [S{\sc ii}], H$\beta$ and [O{\sc iii}], while these lines are not covered by DEIMOS. In total, we have 83 H$\alpha$ emitters in our sample (for 6 we have measurements from both DEIMOS and AF2). Out of these, 75 H$\alpha$ emitters are all at the redshift of the cluster ($z=0.18-0.197$), but some are far from the centre: 52 H$\alpha$ emitters are found within a (projected) radius of 2.5\,Mpc from the cluster ``centre", defined as in \cite{Stroe15} (RA[J2000] 22:42:45.6, Dec[J2000] +53:03:10.8), while 44 are within a 2\,Mpc (projected) radius, 36 are within a 1.5\,Mpc (projected) radius.

%
%
%
%
\begin{table*}
\begin{center}
{\scriptsize
{\centerline{\sc Table 1: H$\alpha$ emitters in our spectroscopic sample with significant detections of at least two emission lines.}}
\begin{tabular}{lccccccccccccc}
\hline
\noalign{\smallskip}
ID    & $\alpha_{\rm J2000}$ & $\delta_{\rm J2000}$   & $z_{spec}$ & $I_{\rm AB}$ & F$_{\rm H\alpha}$ & {\sc fwhm}    & [N{\sc ii}]/H$\alpha$     & [S{\sc ii}]/[S{\sc ii}]     &  [O{\sc iii}]/H$\beta$ & [S{\sc ii}]/H$\alpha$   & Mass  & C  & AGN \\
            &               &                   &        &            &      log$_{10}$         &     km/s           &      &  &        &      & M$_{\odot}$  & dist.  &            \\
\hline
  SSSD-02 & 22:42:51.27 & +52:54:22.07 & 0.1838 & 18.3 & -15.1 & 198 & 0.71 & 1.3 & --- & 0.3  & 10.3 & 2 & --- \\
  SSSD-04 & 22:43:01.57 & +52:55:27.44 & 0.1895 & 19.3 & -15.7 & 50 & 1.02 & 0.2 & --- & ---  & 9.6 & 2 & 1 \\
  SSSD-06 & 22:42:45.98 & +52:56:15.43 & 0.1910 & 20.2 & -15.0 & 102 & 0.07 & 1.8 & 4.2 & 0.2  & 8.5 & 1.5 &  0 \\
  SSSD-07 & 22:42:51.75 & +52:56:28.75 & 0.2291 & 21.2 & -16.0 & 136 & 0.28 & 3.4 & 0.9 & 0.5  & 8.7 & 10 &  0 \\
  SSSD-08 & 22:42:17.30 & +52:57:11.81 & 0.1844 & 21.4 & -16.1 & 104 & 0.31 & 1.9 & 2.4 & 0.2  & 8.7 & 2 &  --- \\

\hline
\end{tabular}
}
\caption{Notes:  Here we show just the five first entries: the full catalogue is published in the on-line version of the paper. The C column indicates the environment/sub-sample of each source (distance from cluster center, Mpc), with sources flagged as 10 being outside the cluster. The AGN column distinguishes between likely AGN which present narrow-lines (1; NLA), broad lines (2; BLA), likely star-forming galaxy (0; SFG) and unclassified (---; UNC).}\label{table:Gal_props}
\end{center}
\end{table*}

\subsection{Multiband photometry and stellar masses}

We use multi-band catalogues derived in \cite{Stroe15} to obtain information on all the emitters and here we explore the $BgVrIz$ photometry to compute stellar masses by spectral energy distribution (SED) fitting. All the photometry is corrected for Galactic extinction \citep[see details in][]{Stroe15}. We use the spectroscopic redshift of each source, but using $z=0.19$ for all sources does not significantly change any of the results. We compute stellar masses for all candidate H$\alpha$ emitters, regardless of having been targeted spectroscopically or not, so we can compare our spectroscopic sample with the full parent sample. The full sample is explored in \cite{Stroe15}.

Stellar masses are obtained by SED fitting of stellar population synthesis models to $BgVrIz$, following \cite{Sobral11,Sobral14}. The SED templates are generated with the \cite{BC03} package using \cite{B07} models, a \cite{Chabrier} IMF, and exponentially declining star formation histories with the form $e^{-t/\tau}$, with $\tau$ in the range 0.1 Gyrs to 10 Gyrs. The SEDs were generated for a logarithmic grid of 200 ages (from 0.1 Myr to the maximum age at $z=0.19$). Dust extinction was applied to the templates using the \cite{Calzetti} law with $E(B-V)$ in the range 0 to 0.5 (in steps of 0.05), roughly corresponding to A$_{\rm H\alpha}\sim0-2$. The models are generated with five different metallicities ($Z=0.0001-0.05$), including solar ($Z=0.02$). Here we use the best-fit template to obtain our estimate of stellar mass, but we also compute the median stellar mass across all solutions in the entire multi-dimensional parameter space for each source, which lie within 1$\sigma$ of the best-fit and thus also obtain the median mass of the 1$\sigma$ best-fits.

%
\begin{figure}
  \centering
 \includegraphics[width=8.2cm]{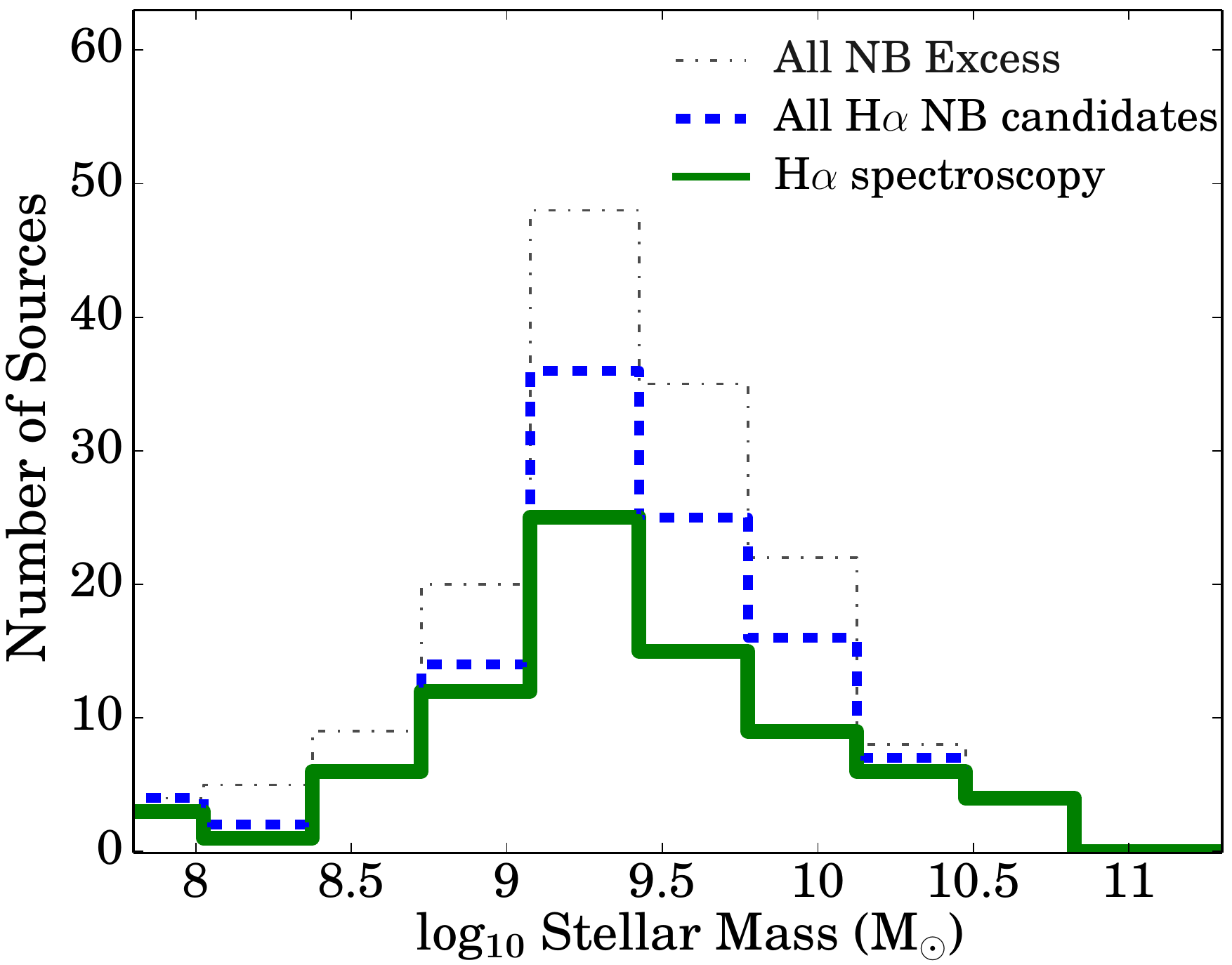}
\caption{Stellar mass distribution for our full spectroscopic sample and comparison with the parent NB sample of all emitters (``All NB Excess''; stellar masses computed assuming all would be at $z=0.19$ and be H$\alpha$ emitters, just shown for comparison, as many sources here are clearly not H$\alpha$ emitters) and the sample of H$\alpha$ emitters at $z=0.19$ after colour-colour, photometric redshift and spectroscopic redshift selection (``All H$\alpha$ NB candidate''). This shows that we are almost fully complete at both low and high masses (compared to the parent sample). Even at intermediate to high masses, where the number of sources is higher, we still have a very high spectroscopic completeness of $\sim50$\% or more. Most importantly, the sources that are not in our spectroscopic sample are those that i) we could not target due to fibre configuration constraints and ii) that have very low fluxes.}
\label{Mass_distri}
\end{figure}

\subsection{Completeness: Stellar Mass}

We show the distribution of stellar masses for the samples of H$\alpha$ emitters in Figure \ref{Mass_distri}. H$\alpha$ emitters in our full sample have an average stellar mass of $\sim10^{9.4}$\,M$_{\odot}$. As a whole, the sample of H$\alpha$ emitters at $z\sim0.19$ has a similar stellar mass distribution to samples of field H$\alpha$ emitters at similar redshifts \citep[e.g.][]{Shioya, Sobral14}, but with cluster H$\alpha$ emitters having higher stellar masses than H$\alpha$ emitters outside the cluster. Figure \ref{Mass_distri} also shows that our main limitation at lower masses is our parent sample from \cite{Stroe15}, which is complete down to roughly $\sim10^{9}$\,M$_{\odot}$, and thus our results, particularly for the mass-metallicity relation, only take into account star-forming galaxies with stellar masses $>10^{9}$\,M$_{\odot}$.

%
%
%
\begin{figure}
  \centering
 \includegraphics[width=8.2cm]{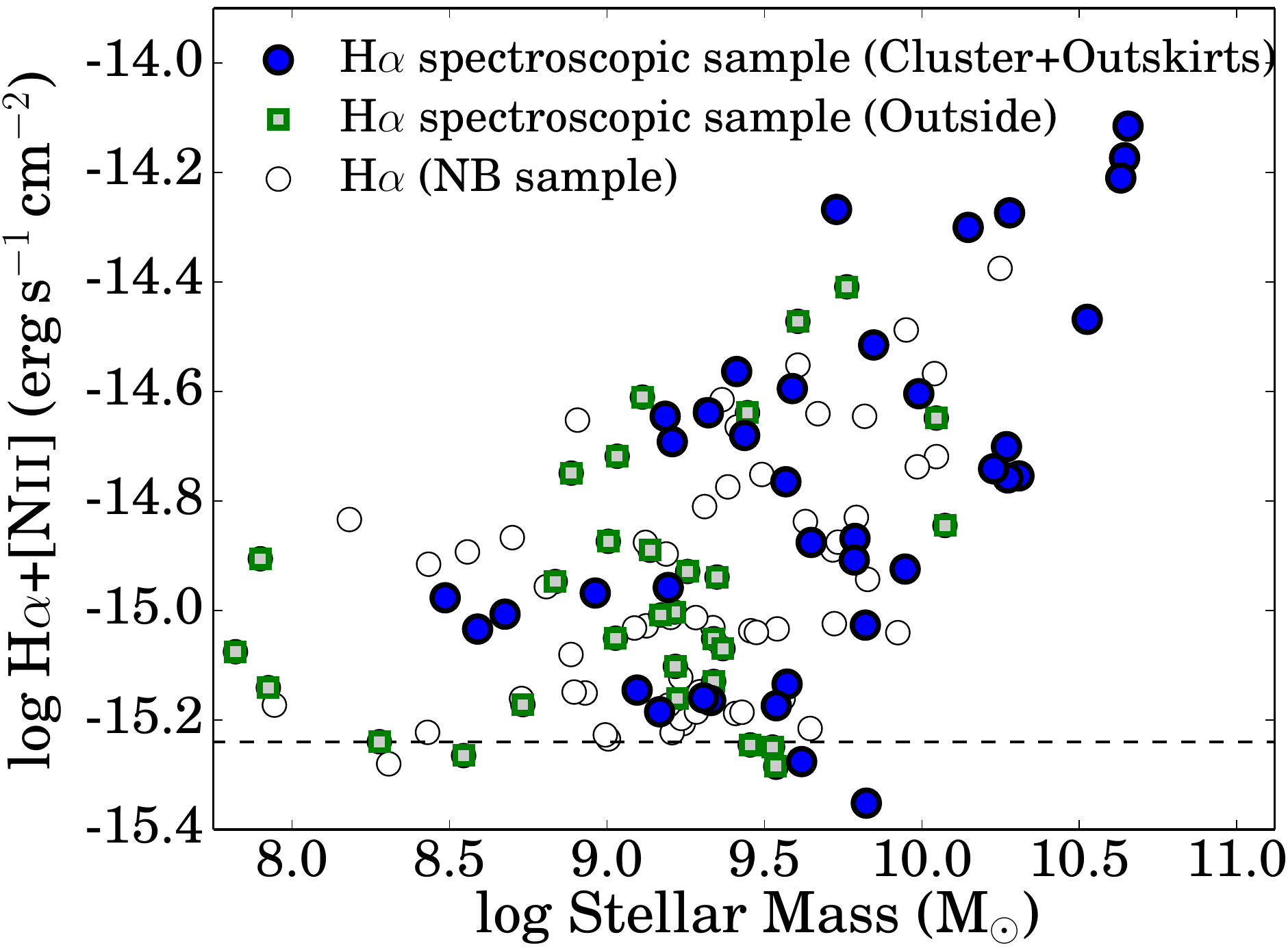}
\caption{The relation between H$\alpha$+[N{\sc ii}] flux (corrected for Galactic extinction, but not corrected for intrinsic dust extinction), based on 5$''$ narrow-band photometry and stellar mass, for the parent sample, selected using narrow-band, and our spectroscopic sample. Our spectroscopically confirmed sources sample the vast majority of the parameter space, both for galaxies in the cluster ($R<1.5$\,Mpc) and outskirts ($1.5<R<2.5$\,Mpc) and for those outside the cluster. We are particularly complete, relative to the parent sample, for stellar masses $>10^{9}$\,M$_{\odot}$. We preferentially miss sources with the lowest fluxes and with stellar masses lower than $\sim10^{9}$\,M$_{\odot}$.}
\label{SFR_vs_MASS}
\end{figure}

\subsection{Completeness: SFR-Stellar Mass}

Figure \ref{SFR_vs_MASS} shows the relation between H$\alpha$(+[N{\sc ii}]) flux (based on narrow-band photometry, so we can fully compare it with the parent NB sample) and stellar mass, for both the parent sample, and for our spectroscopic sample. We also highlight sources confirmed to be outside the cluster, and those in the cluster and outskirts. The comparison with the parent sample shows that our sample is representative of the full parent sample, at least down to stellar masses of $>10^{9}$\,M$_{\odot}$, and for fluxes (corrected for Galactic extinction and for 5$''$ apertures, and thus in practice after a full aperture correction) of H$\alpha$ flux $>10^{-15.25}$\,erg\,s$^{-1}$\,cm$^{-2}$ (roughly corresponding to SFRs$>0.2$\,M$_{\odot}$\,yr$^{-1}$).

\subsection{The comparison sample: DEIMOS+AF2 non-cluster H$\alpha$ emitters} \label{OTHER_comparison}

We explore our H$\alpha$ emitters in the DEIMOS dataset (8) that are found to be at higher ($0.23<z<0.3$) and lower redshift ($0.14<z<0.17$) , and 31 H$\alpha$ emitters from the AF2 data-set that are more than 2.5\,Mpc away from the Sausage cluster ``centre'' but at a similar redshift. As mentioned in \S\ref{Redshifts}, we use RA(J2000) 22:42:45.6, Dec(J2000) +53:03:10.8 as the ``centre" of the Sausage merging cluster, and compute projected distances from this position. This sample of 39 H$\alpha$ emitters is compared with a similar number of Sausage cluster H$\alpha$ emitters and allows us to directly compare their properties, AGN contamination and search for any differences. We use this sample for direct comparisons.

%
%
%
%
\begin{figure}
  \centering
 \includegraphics[width=8.2cm]{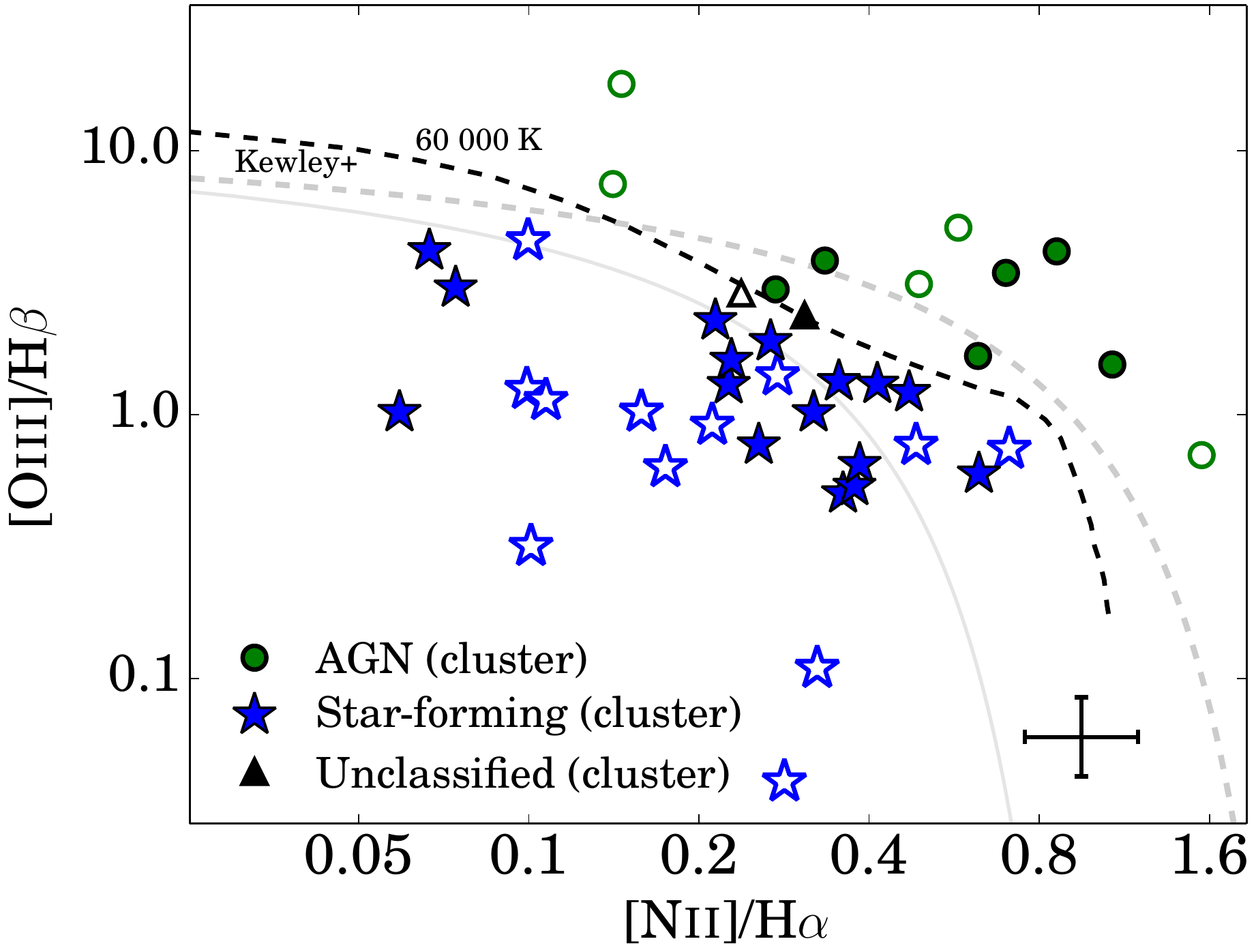}
\caption{Emission line ratio diagnostics \citep[][]{BPT,Rola} separate star-forming dominated from AGN dominated H$\alpha$ emitters (black dashed line). We also show emission line diagnostics from \citet{Kewley01} for comparison, which show the location of pure, ``typical'' star-forming galaxies (gray solid line), and the separation line between maximal starbursts and AGN (gray dashed line). We only show galaxies with detections in all emission lines. Filled symbols are H$\alpha$ emitters within a 1.5\,Mpc radius of the cluster, while the unfilled symbols are either at higher, lower redshift, or are at the redshift of the cluster, but at distances higher than 1.5\,Mpc. These results reveal a similar fraction of AGN in ($36\pm8$\%) and outside ($29\pm7$\%) the cluster. Note that, due to the significant dust extinction, particularly on the line of sight, the [O{\sc iii}]/H$\beta$ line ratio is slightly overestimated for all galaxies (likely by $\sim0.06$\,dex), making it easier to classify galaxies as AGN, and making the sample of star-forming galaxies even cleaner from potential AGN contamination.}
\label{AGNvsSF}
\end{figure}

%
%
\begin{figure*}
  \centering
 \includegraphics[width=16.2cm]{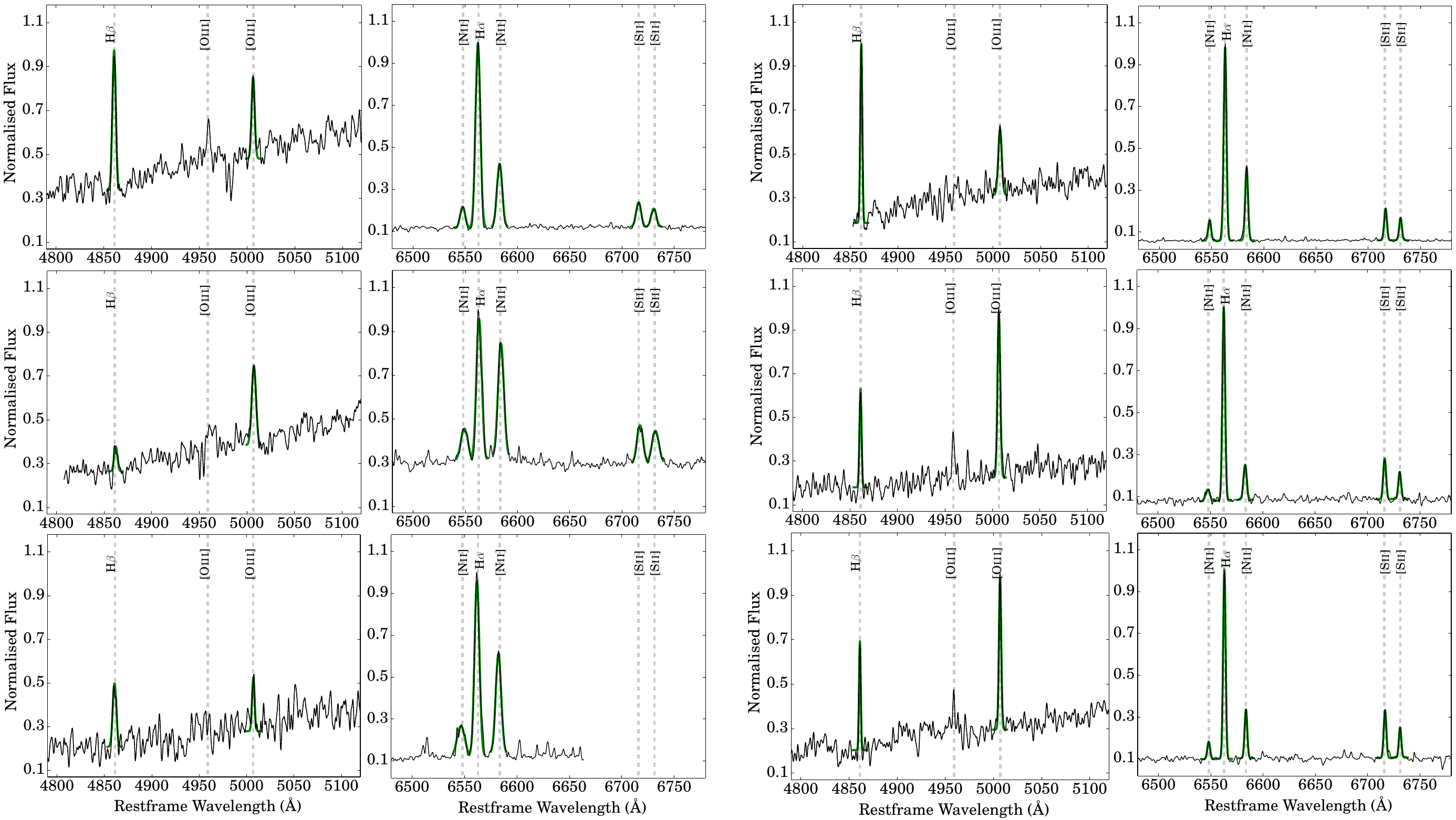}
\caption{Some examples of our H$\alpha$ emitters, the coverage that extends to H$\beta$ and [OIII] and the gaussian fits we derived in order to measure emission line ratios. This extended coverage is particularly important in order to allow us to distinguish between star-forming dominated and AGN-dominated sources by measuring [O{\sc iii}]\,5007/H$\beta$ and [N{\sc ii}]\,/H$\alpha$ line ratios and placing them on the \citealt{BPT} classification scheme.}
\label{SPECTRA_HBOIIIHA}
\end{figure*}

\section{Results} \label{Results}

The redshift distribution of our final sample of H$\alpha$ emitters belonging to the merging cluster is shown in Figure \ref{zspec_dist} and compared to the distribution of spectroscopic redshifts for H$\alpha$ emitters within different radii from the cluster centre. We confirm that the narrow-band filter used in \cite{Stroe15} effectively selects all H$\alpha$ emitters belonging to the merging cluster, and that such selection is not biased towards galaxies at the outskirts (in the redshift direction) of the cluster. We fully confirm the very high number of H$\alpha$ emitters in this merging cluster. Given the mass of the cluster \citep[$\sim 1.0-1.1\times 10^{15} M_\odot$, see][]{Jee2015}, and its very high ICM temperature \citep[][Akamatsu et al. 2015]{Ogrean13} -- 7-12 KeV --, it is puzzling that there are so many H$\alpha$ emitters. With a final sample of 39 field H$\alpha$ emitters and samples of 52, 44 and 36 H$\alpha$ emitters (within 2.5, 2.0 and 1.5 \,Mpc from the cluster ``centre", respectively) we now investigate their nature and unveil their properties. For the remaining analysis in the paper, we divide our sample in three different environments: i) Cluster (sources at the redshift of the cluster and within R$<1.5$\,Mpc), ii) Cluster outskirts (sources at the redshift of the cluster and at projected distanced  $1.5<R<2.5$\,Mpc) and iii) Outside the cluster (sources at the redshift of the cluster that are found to be $R>2.5$\,Mpc away and sources at a significantly higher and lower redshift). For some parts of the analysis, we also split the cluster sample into galaxies in the i) cluster, near to the hottest intra-cluster medium, $R<0.5$\,Mpc away from the ``centre" of the cluster and ii) post-shock region, within the North and South radio relics, close to the relics and further away from the ``centre". We refer to ii) as ``post-shock" region and to i) as "elsewhere in the cluster''.

\subsection{Nature of H$\alpha$ emitters: SF vs AGN} \label{SF_AGN}

In order to differentiate between star-forming and AGN, the [O{\sc iii}]\,5007/H$\beta$ and [N{\sc ii}]\,5007/H$\alpha$ line ratios were used (see Figure \ref{AGNvsSF}); these have been widely used to separate AGN from star-forming galaxies \citep[e.g.][]{BPT,Rola,Kewley01,Kewley13}. We show some examples of spectra in Figure \ref{SPECTRA_HBOIIIHA}. These line ratios are also for emission lines sufficiently close that dust extinction has little effect. However, for the case of the [O{\sc iii}]\,5007/H$\beta$ emission line ratios, due to the significant total dust extinction affecting our galaxies, particularly due to the Galaxy, line ratios may be over-estimated by $\sim0.06$\,dex. Because we do not correct for this, the [O{\sc iii}]\,5007/H$\beta$ line ratios are all closer to AGN. This means, however, that our sample of star-forming galaxies will be even more conservative and robust (if anything, some star-forming galaxies may be classified as AGN). Because corrections are relatively unreliable, and because applying unreliable corrections could lead to including potential AGN in our samples of star-forming galaxies, we opted not to correct for this effect. Only spectra with all lines detected at S/N\,$>3.0$ were used, but we also place limits on those with lower S/N. Figure \ref{AGNvsSF} shows data-points for the line ratios, while the black dashed curve shown represent maximum line ratios for a star-forming galaxy \citep[from OB stars with effective temperatures of 60000\,K;][]{BPT,Rola}. We also show curves from \cite{Kewley01} and \cite{Kewley13} encompassing ``pure'', ``typical'' star-forming galaxies (gray solid line), and encompassing up to maximal starbursts (gray dashed line).

Over our full AF2 and DEIMOS sample, we find 4 broad line AGNs. All these broad line AGNs are found to be in the cluster. Furthermore, in total, we have measurements of [O{\sc iii}]\,5007/H$\beta$ and [N{\sc ii}]\,/H$\alpha$ line ratios with individual line detections above 3\,$\sigma$ which allow us to distinguish between AGN and SF for 42 sources. For these 42 sources, we find 14 AGN (10 narrow-line AGN and 4 BL-AGN), and 28 likely star-forming dominated H$\alpha$ emitters. We show the location of these sources in Figure \ref{SAUSAGE_SKY}, revealing that AGN in the cluster are all in the post-shock regions, just behind both the North and South radio relics/shock fronts.

\begin{figure}
  \centering
 \includegraphics[width=8.2cm]{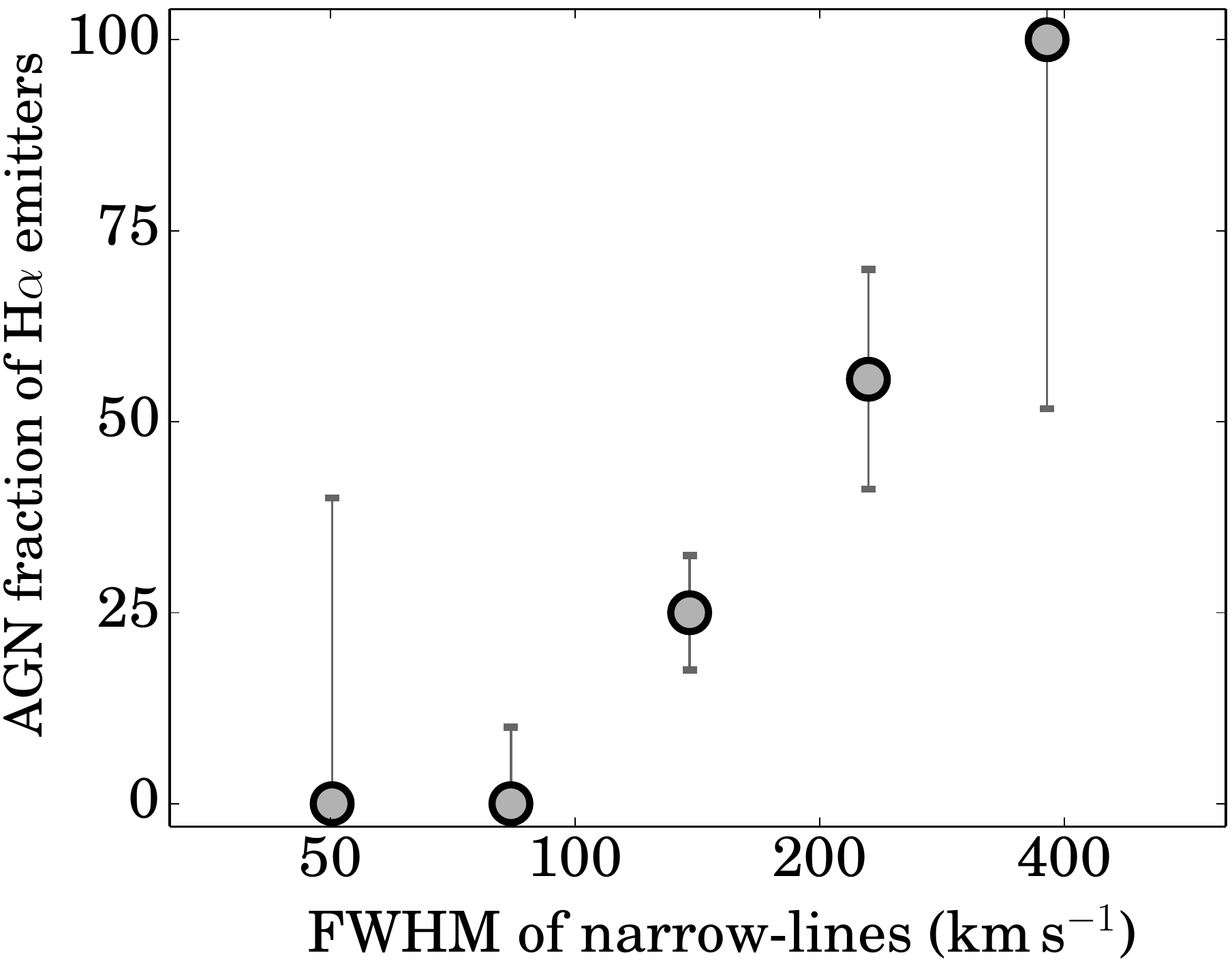}
\caption{Fraction of AGN-dominated galaxies as a function of emission line FWHM in km\,s$^{-1}$ for the narrow lines (broad emission lines are neglected here). We find that at higher FWHM of H$\alpha$, [N{\sc ii}], [S{\sc ii}] lines, the prevalence of AGN increases, likely indicating that AGN are the cause for such high FWHM in narrow lines, and indicative of outflows.}
\label{FWHM_AGN}
\end{figure}

For the sources we can classify we also have measured the FWHM of the narrow emission lines. We show the fraction of AGN sources as a function of FWHM of the narrow lines in Figure \ref{FWHM_AGN}. This clearly shows that at the highest FWHM, the AGN fraction is very high. We note that these are FWHM of narrow lines, and thus this is likely indicative of outflows happening in the AGN in our sample, dominated by those in the Sausage cluster.

We split sources between those in the cluster (see Figure \ref{SAUSAGE_SKY}) and outskirts (25 classified sources) and those outside the cluster (17 classified sources). We find 9/25 sources in the cluster+outskirts to be AGN (including the 4 broad-line AGN), resulting in an AGN fraction of $36\pm8$\,\% (Poissonian errors), while outside the cluster we find 5/17 sources to be AGN, resulting in an AGN fraction of $29\pm7$\,\%, lower than in the cluster, but still consistent. It should be noted that both samples have very similar median H$\alpha$ luminosities, and thus should be fully comparable. For H$\alpha$ emitters within the Sausage merging cluster (R$<1.5$\,Mpc), we find an AGN fraction of $35\pm6$\% (see Figure \ref{SAUSAGE_SKY}).

%
%
%
%

%
%
%
%
\begin{figure*}
  \centering
 \includegraphics[width=14cm]{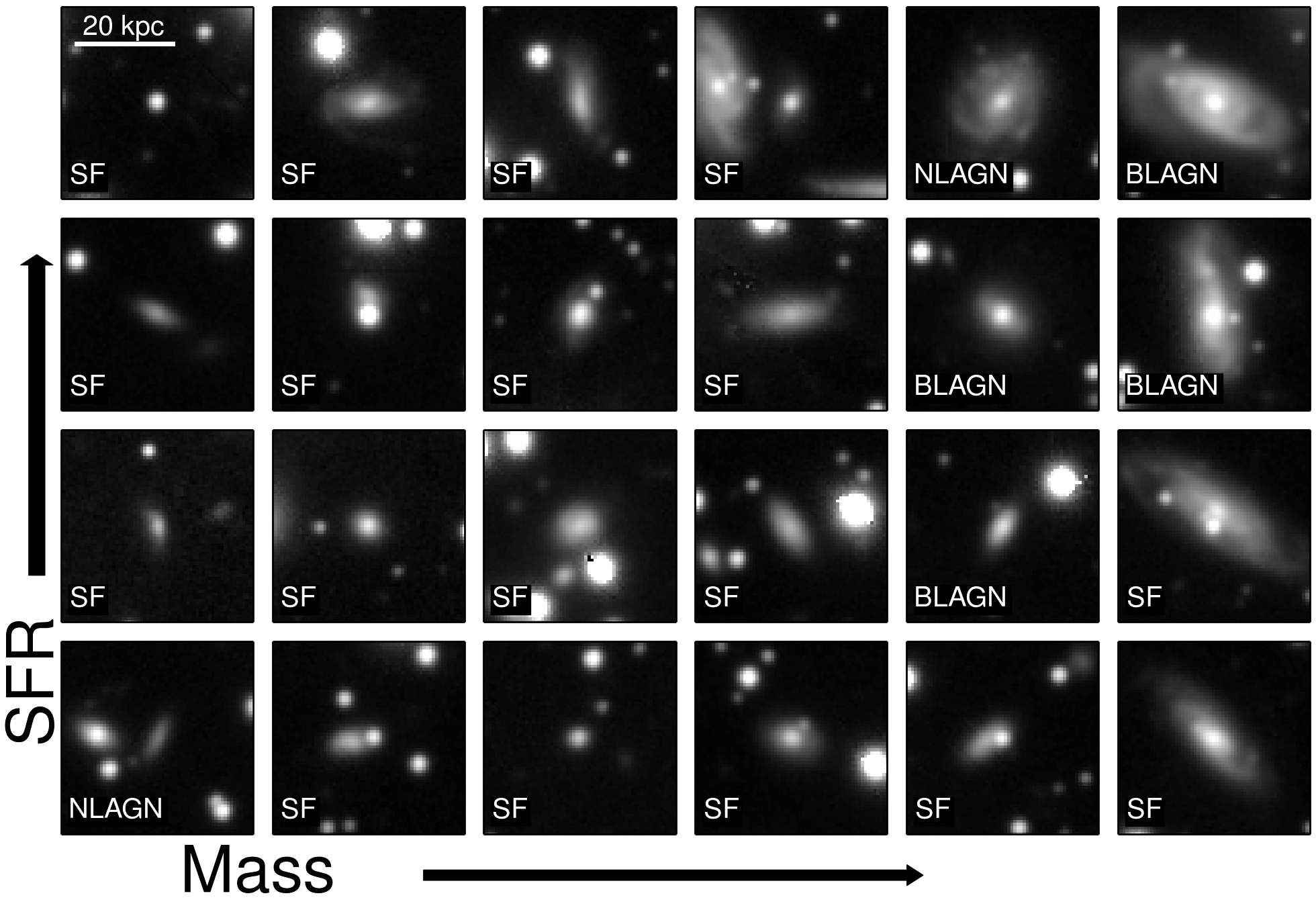}
\caption{Thumbnails of our H$\alpha$ cluster members. Each square is $\sim40\times40$\,kpc and we organise galaxies in respect to their estimated stellar mass and estimated star-formation rate. We also indicate which sources are likely AGN and which are star-forming. We do not find any evidence for significant galaxy-galaxy (major) merging. This implies that the enhanced star-formation and AGN activity within the merging cluster is not being driven by galaxy-galaxy mergers, and thus is it more likely driven by the interaction with the environment, and in particular with the shock wave. We note that the bright point sources, which show up in most images are stars within the Milky Way, not galaxies.}
\label{Morphologies}
\end{figure*}

\subsection{Morphologies}\label{Morphologies}

By exploring deep $i$ band Subaru images (see also \cite{Stroe14a}), we investigate the morphologies of our H$\alpha$ emitters. We show thumbnails of all our H$\alpha$ cluster galaxies, also labelling them as AGN or star-forming galaxies, in Figure \ref{Morphologies}. We find little to no indication of merger activity (note that the stellar density from the Galaxy is extremely high: point-like sources are stars). We however note that most star-forming galaxies show relatively compact morphologies and hint that most star-formation is occurring in relatively central regions, where molecular gas is likely still available to form stars. However, a more detailed morphological analysis is beyond the scope of this paper.

Field star-forming galaxies at these H$\alpha$ luminosities present a typical fraction of mergers on the order of $\sim10$\% \citep[e.g. at $z=0.24$ in the COSMOS field;][]{Sobral09}, and our H$\alpha$ emitters in the cluster do not present a larger fraction than that. Thus, the elevated activity in our cluster H$\alpha$ emitters is definitely not being driven by mergers as, if anything, our H$\alpha$ emitters have a lower fraction of mergers than those in the field. This is, nonetheless, not surprising. The cluster we are studying is incredibly massive, with a high velocity dispersion of over 1000\,km\,s$^{-1}$, and thus the chances of a galaxy-galaxy mergers are relatively small.

\subsection{Electron densities and Ionisation Potential} \label{Elec_dens_IONISA}

We make clear individual detections of the [S{\sc ii}]$_{6716,6761}$ doublet.  We also (median) stack the entire sample to find [S{\sc ii}]$_{6716}$/ [S{\sc ii}]$_{6761}$= 1.22 $\pm$ 0.05, corresponding to an electron density of $10^{2.4\pm0.1}$\,cm$^{-3}$\citep[][]{Osterbrock89}. If we only consider star-forming galaxies, we find [S{\sc ii}]$_{6716}$/ [S{\sc ii}]$_{6761}$= 1.36 $\pm$ 0.07, corresponding to $10^{2.00\pm0.25}$\,cm$^{-3}$.

%
%
%
%
\begin{figure*}
  \centering
 \includegraphics[width=14.8cm]{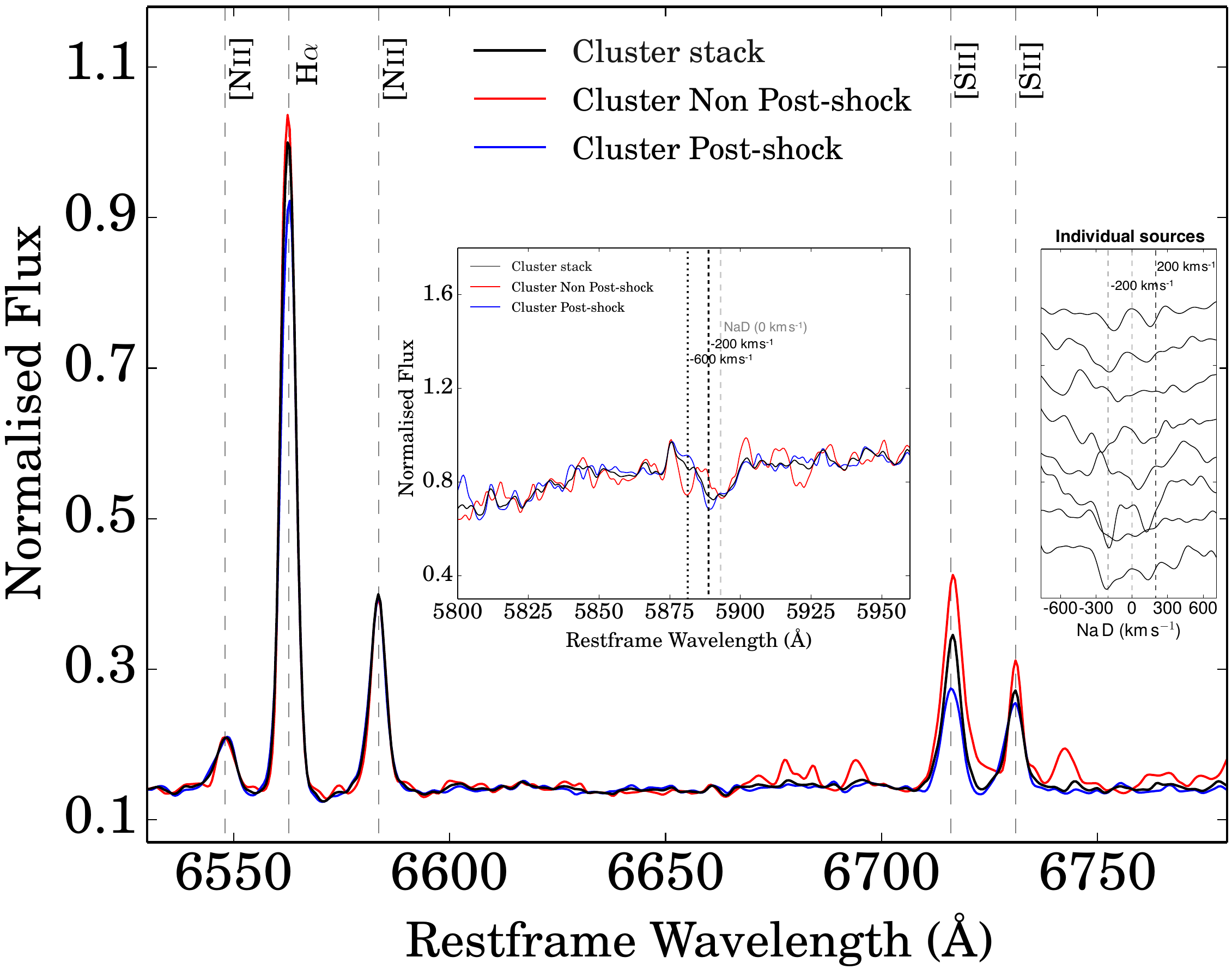}
\caption{Stacks for our full merging cluster star-forming galaxies, and when splitting the sample in post-shock regions and non post-shock regions (mostly those in the very hot intra-cluster medium and in the sub-cluster cores). While as a whole cluster galaxies show significant differences from field and outskirts star-forming galaxies, post-shock and non post-shock star-forming galaxies also show a drastic difference in their stacked spectra regarding the [S{\sc ii}] emission lines, and particularly for [S{\sc ii}]$_{6716}$, which is significantly boosted for non post-shock cluster star-forming galaxies. This may be evidence of significant supernova activity. We also find a significant red-shifted component of the [S{\sc ii}]$_{6716}$ emission line, potentially indicative of outflows, and is relatively broad. Further evidence for outflows is found even for the stack of both sub-samples: we find NaD absorption line significantly blue-shifted, from 200 to 600\,km\,s$^{-1}$, a clear sign that, as a whole, cluster star-forming galaxies are driving rapid outflows \citep{Heckman2000}. We note that Na\,D in absorption with significant velocity offset from the systematic redshift within the range 200 to 600\,km\,s$^{-1}$ is also found for individual sources with even stronger S/N (as the stack mixes different velocity offsets); we show those in the right panel.}
\label{STACKS_and_outflows}
\end{figure*}

In order to compare several sub-samples, based on membership and nature, we further split the sample in i) all (all sources), ii) cluster, iii) outskirts and iv) outside. Within the cluster sample, we further split it into sources within the post-shock regions (both North and South, see Figure \ref{SAUSAGE_SKY}), and those elsewhere, including in the two cluster cores and within the hot X-ray gas. We further split samples with respect to the dominating nature of the sources: i) all sources, ii) sources clearly dominated by star formation (SFGs) and iii) sources dominated by AGN activity (AGNs). The results are presented in Table \ref{table:RESULTS}.

Our results clearly show that all sub-samples of Sausage cluster members have higher [S{\sc ii}]$_{6716}$/ [S{\sc ii}]$_{6761}$ line ratios than similar sub-samples. In particular, the merging cluster star-forming galaxies show a very high [S{\sc ii}]$_{6716}$/ [S{\sc ii}]$_{6761}$= 1.73 $\pm$ 0.11, implying an extremely low electron density of $<5$\,cm$^{-3}$ \citep[the higher the line ratio, the lower the electron density;][]{Osterbrock89}, $<30$ times lower electron density than star-forming galaxies outside the cluster and other star-forming galaxies found in the literature. On the other hand, it should be noted that AGNs (see Table \ref{table:RESULTS}) all have [S{\sc ii}]$_{6716}$/ [S{\sc ii}]$_{6761}$ line ratios below 1, as expected, so completely opposite to what is found for the cluster star-forming galaxies. 

By further splitting the cluster sample into galaxies in the post-shock region (within the North and South radio relics, close to the relics and the furthest away from the ``centre'') and those elsewhere, and particularly in the hot intra-cluster medium (see Table \ref{table:RESULTS}), we show that the high [S{\sc ii}]$_{6716}$/ [S{\sc ii}]$_{6761}$ line ratio is being strongly driven by star-forming galaxies within the hottest inter-cluster medium (those closest to the ``centre", and further away from the shock fronts). These star-forming galaxies (no AGNs are found, but 2 are unclassified) show very high [S{\sc ii}]$_{6716}$/ [S{\sc ii}]$_{6761}$ = $2.5\pm0.2$, corresponding to extremely low electron densities. This is likely evidence that such star-forming galaxies are substantially affected by their surrounding environment. Most importantly, the stack of the non-post-shock galaxies (see Figure \ref{STACKS_and_outflows}) reveals asymmetric [S{\sc ii}]$_{6716}$ emission line, with significantly blue-shifted emission, likely indicating stripping/outflows.

%
%
%
%
%
%
\begin{table*}
\begin{center}
{\scriptsize
{\centerline{\sc Results from the stacks for different sub-samples.}}
\begin{tabular}{cccccc}
\hline
\noalign{\smallskip}
SAMPLE   & number sources & [N{\sc ii}]/H$\alpha$     & 12\,+\,log(O/H) & [S{\sc ii}]/[S{\sc ii}] &  [S{\sc ii}]/H$\alpha$   \\
\hline
  Full Sample & 83 & $0.338\pm0.007$ & $8.632\pm0.005$ & $1.22\pm0.05$ & $0.249\pm0.003$ \\
  \bf All in Cluster & 24 & $0.443\pm0.007$ & $8.698\pm0.004$ & $1.48\pm0.08$  &  $0.285\pm0.002$    \\
  \it All in Post-shock (PS) & 17 & $0.655\pm0.01$ & $8.795\pm0.004$ & $1.24\pm0.06$ & $0.32\pm0.003$ \\
  \it All Cluster non-PS & 7 & $0.265\pm0.005$ & $8.571\pm0.005$ & $2.23\pm0.18$ & $0.372\pm0.004$ \\
  All in Outskirts & 20 & $0.287\pm0.007$ & $8.591\pm0.006$ & $0.96\pm0.04$ &   $0.209\pm0.004$    \\
  All Outside& 31 & $0.168\pm0.006$ & $8.458\pm0.009$ & $1.26\pm0.04$  &  $0.234\pm0.005$   \\
  \hline
  All SFGs & 28 & $0.284\pm0.006$ & $8.588\pm0.005$ & $1.43\pm0.07$   &  $0.259\pm0.002$  \\
   \bf Cluster SFGs & 11 & $0.311\pm0.005$ & $8.611\pm0.004$ & $1.73\pm0.11$ & $0.265\pm0.002$  \\
   \it Post-shock (PS) SFGs & 6 &  $0.339\pm0.005$ & $8.632\pm0.004$ & $1.22\pm0.04$ & $0.201\pm0.002$ \\
   \it Cluster non-PS SFGs & 5 & $0.285\pm0.005$ & $8.590\pm0.004$ & $2.5\pm0.2$ & $0.430\pm0.004$ \\
   Outskirts SFGs & 5 & $0.180\pm0.002$ & $8.476\pm0.003$ & $1.57\pm0.10$ & $0.193\pm 0.002$    \\
  Outside SFGs & 10 & $0.152\pm0.010$ & $8.433\pm0.016$ & $0.82\pm0.04$ & $0.212\pm0.007$ \\
  \hline
  All AGNs & 17 & $0.731\pm0.009$ & --- & $0.93\pm0.02$ &  $0.250\pm0.003$\\
  \bf  Cluster AGNs & 6 & $0.938\pm0.014$ & --- & $0.82\pm0.02$  &    $0.251\pm0.003$  \\
  \it Post-shock (PS) SFGs & 6 & $0.938\pm0.014$ & --- & $0.82\pm0.02$  &    $0.251\pm0.003$  \\
   \it Cluster non-PS SFGs & 0 & --- & --- & --- & --- \\
  Outskirts AGNs & 5 & $1.054\pm0.031$ & --- & $0.58\pm0.02$ & $1.1\pm0.1$ \\
  Outside AGNs & 7 & $0.513\pm0.007$ & --- & $0.62\pm0.03$ & $0.13\pm0.01$ \\
\hline
\end{tabular}
}
\caption{Notes: The Full sample contains all H$\alpha$ emitters, the Sausage Cluster sample is defined with sources within 1.5\,Mpc radius of what we assign as the central position of the cluster. The sample in the outskirts is defined as H$\alpha$ emitters within 1.5 and 2.5 projected Mpc from the central position, and sources defined as outside are at higher distances than 2.5 Mpc projected.} \label{table:RESULTS}
\end{center}
\end{table*}

The [S{\sc ii}]$_{6716}$/H$\alpha$ line ratio can be used to estimate the ionisation strength \citep[][]{Osterbrock89,Collins2001} of the inter-stellar medium (ISM). We derive, for our full sample (median stack), [S{\sc ii}]$_{6716}$/H$\alpha$ = 0.249 $\pm$ 0.003 (see Table \ref{table:RESULTS}), which corresponds to an ionisation parameter log$_{10}$(U, cm$^3$) = $-4.06 \pm 0.05$ \citep{Collins2001}. Cluster members show the highest [S{\sc ii}]$_{6716}$/H$\alpha$ ratios. Focusing on the H$\alpha$ star-forming galaxies in the Sausage merging cluster, we find [S{\sc ii}]$_{6716}$/H$\alpha$ = 0.265 $\pm$ 0.002, which corresponds to a ionisation strength of the ISM about half of that of the field and outskirts sample. However, H$\alpha$ star-forming galaxies in the outskirts and outside the cluster are significantly more metal poor (see \S\ref{metal}), which is enough to explain the difference. When matched in metallicities, we find no significant difference within the errors.

When we further split the cluster sample into sources in the post-shock region and those elsewhere (mostly in the hottest X-ray gas region, near the ``centre" of the cluster and/or in the sub-cluster cores), we find that the high [S{\sc ii}]$_{6716}$/H$\alpha$ ratio within the cluster is mostly driven by cluster star-forming galaxies outside the post-shock region, again indicating that these galaxies are affected by their surroundings. In practice, with a [S{\sc ii}]$_{6716}$/H$\alpha$= $0.430\pm0.004$, cluster star-forming galaxies away from the post-shock regions have an ionisation parameter log$_{10}$(U, cm$^3$) = $-4.5 \pm 0.05$, more than 4 times lower than all other star-forming galaxies in the cluster. This could be interpreted as further evidence that these galaxies are already having their star-formation activity quenched. However, we note that this very high [S{\sc ii}]$_{6716}$/H$\alpha$ ratio could also be interpreted as a significant contribution from supernova remnants. Since we do not find any difference in the typical SFRs of these galaxies relative to the other star-forming galaxies in and outside the cluster, the supernova explanation is strongly favoured. Furthermore, as we find evidence for outflows (see Figure \ref{STACKS_and_outflows}), both in redshifted [S{\sc ii}]$_{6716}$ emission, but particularly in strongly blue-shifted Na\,D \citep[see e.g.][]{Heckman2000} absorption ($\sim600$\,km\,$^{-1}$) for these star-forming galaxies, it may well be that these outflows are being driven by supernovae.

\subsection{Outflows}

Particularly focusing on the Keck/DEIMOS sample (where the S/N is the highest, detecting the continuum for the bulk of the sample), we inspect the H$\alpha$, [N{\sc ii}] and [S{\sc ii}] lines to look for asymmetric profiles, broad components (for the forbidden lines) and P Cygni profiles, all potential signatures of strong outflows. We find strong evidence for at least one of such signatures in 7 of our 24 cluster galaxies, while we find no such signatures for galaxies outside the cluster (but the Keck/DEIMOS sample outside is smaller). For the Keck/DEIMOS sample only (as it is the only data-set that actually allows us to detect such signatures at the necessarily high S/N in a complete way), we find such signatures in $\sim22$\,\% of the cluster sample, and 100\% of these are in the post-shock regions (see Figure \ref{SAUSAGE_SKY}). For the 7 sources, the absorption features show offsets of 600-1000\,km/s. Many of these are AGNs and, as discussed in \S \ref{SF_AGN}, all cluster AGNs are in the post shock-front regions of both north and south relics/shock-waves (see Figure \ref{SAUSAGE_SKY}).

We also attempt to fit emission lines with a combination of a narrow and a broad component. Whenever the S/N for the bluer lines (H$\beta$ and [O{\sc iii}]) is lower than 10 we use only H$\alpha$, [N{\sc ii}] and [S{\sc ii}]. We find that a single Gaussian profile (with a FWHM of up to 500\,km\,s$^{-1}$) is able to fully fit all the spectra apart from the BL-AGNs. This also holds true for the stacks. We note that given the lower spectral resolution of WHT/AF2 when compared to Keck/DEIMOS (and lower S/N per \AA, see e.g. Figure \ref{AF2_DEIMOS}), we find that we can only reliably measure FWHM of emission lines with AF2 if they are larger than 160\,km\,s$^{-1}$. With DEIMOS, we can measure FWHM down to 60-80\,km\,s$^{-1}$. For the Keck/DEIMOS sample, we find that that the average FWHM is $156\pm84$\,km\,s$^{-1}$ (for AF2 we find an average of $174\pm70$\,km\,s$^{-1}$). The line ratios and other properties for our full sample are given in Table \ref{table:Gal_props}. We find that the fraction of AGN correlates with increasing FWHM of narrow lines (see Figure \ref{FWHM_AGN}), indicating that AGN are likely driving strong outflows.

Finally, for the sources with the highest S/N in the continuum, for which we can detect clear absorption lines, we also measure systematic velocity offsets from the absorption and emission lines. We find strong evidence for outflows (see Figure \ref{STACKS_and_outflows}), both in redshifted [S{\sc ii}]$_{6716}$ emission, but particularly in strongly blue-shifted Na\,D absorption ($\sim200-600$\,km\,$^{-1}$) for cluster star-forming galaxies as a whole (median stack). We also look at Na\,D in absorption which may be offset significantly on a source by source basis. We do this by fitting Na\,D with a Gaussian profile, and then comparing net velocity offsets when compared to the median redshift given by all the emission lines. We find strong evidence for outflows in all cluster star-forming galaxies except one (see Figure \ref{STACKS_and_outflows}). We find an average velocity offset of 210$\pm70$\,km\,s$^{-1}$, in line with the stack. The most important result is that the vast majority of the Sausage merging cluster star-forming galaxies are driving strong outflows, and thus are experiencing (stellar) feedback. An alternative would be that these galaxies are having their gas stripped into the intra-cluster medium. However, if the latter was the case, one would expect that the velocity offsets would largely average out to zero, since the relative motion of the galaxies with respect to the gas should be random. We therefore argue that it is much more likely that we are witnessing strong stellar feedback which, of course, given the environment, will likely mean all the gas is permanently removed from the galaxies.

%
%
%
%
%
\begin{figure*}
 \centering
  \includegraphics[width=18cm]{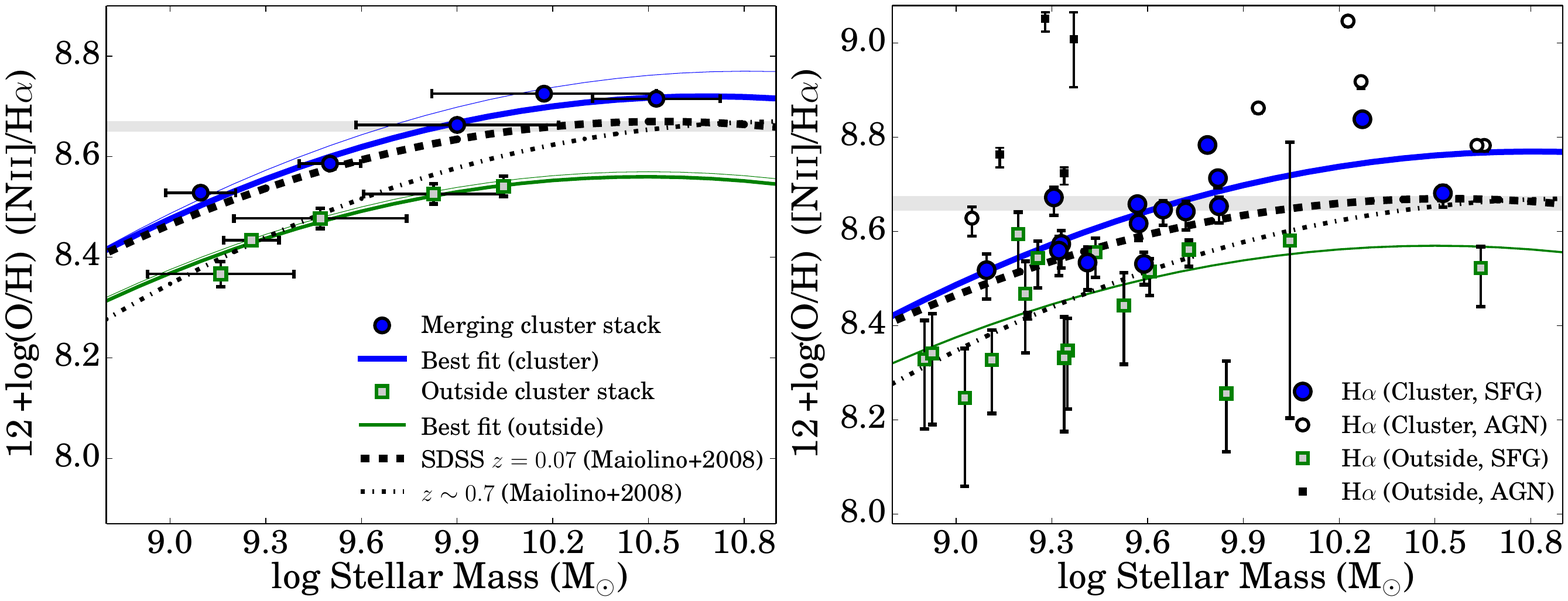}
\caption{The mass metallicity relation for our H$\alpha$ emitters. Grey lines indicate solar metallicity, for reference. We find a good correlation between (stacked, left panel; individual measurements, right panel) metal abundance (excluding all AGN), traced by [N{\sc ii}]/H$\alpha$, and stellar mass of each galaxy, for both H$\alpha$ emitters in the cluster (within a radius of 1.5\,Mpc), and for H$\alpha$ emitters in our comparison sample, outside the cluster, observed with the same instruments and with the same selection functions. H$\alpha$ emitters in the cluster are systematically more metal rich at fixed mass than those in the comparison sample, and follow closely the local SDSS mass-metallicity relation, or even higher, particularly at high masses. On the other hand, H$\alpha$ emitters outside the Sausage cluster reveal some evolution relative to the SDSS $z=0$ relation \citep[][after applying the appropriate corrections for a different metallicity indicator and a different IMF]{Maiolino08}. When directly comparing our sample of H$\alpha$ emitters in the Sausage cluster and those outside the cluster, we find a systematic offset of about 0.2\,dex, which gets tentatively higher for higher stellar masses. This shows that H$\alpha$ emitters in the Sausage are more metal rich.}
\label{Mass_metallicity}
\end{figure*}

\subsection{Metallicities} \label{metal}

We use the [N{\sc ii}]/H$\alpha$ emission line ratio to infer the metallicity of the gas for each star-forming galaxy (AGNs are neglected). We obtain metallicities for each star-forming source, but also for sub-samples: see Table \ref{table:RESULTS}. For our full sample (median stack), we find [N{\sc ii}]\,/\,H$\alpha$\,=\,$0.338\pm0.007$. The [N{\sc ii}]/H$\alpha$ line ratio can be used to obtain the metallicity of our star-forming galaxies (oxygen abundance), [12\,+\,log(O/H)], by using the conversion of \cite{Pettini04}: 12\,+\,log(O/H)\,=\,8.9\,+\,0.57\,log([N{\sc ii}]\,/\,H$\alpha$). The galaxies in our full sample (without excluding AGN) have a median metallicity of $8.632\pm0.005$, which is consistent with solar (8.66$\pm$0.05), but we note that we are sampling galaxies with a range of masses, and thus we need to take that into account when properly comparing the samples -- this is done in \S\ref{Mass_metal}.

Our results reveal that cluster H$\alpha$ emitters have the highest [N{\sc ii}]\,/\,H$\alpha$\,=\,$0.443\pm0.007$ line ratios. However, AGN typically have high [N{\sc ii}]\,/\,H$\alpha$ line ratios, and it is mandatory to exclude them if metallicities are to be robustly estimated from this line ratio. Nevertheless, even when considering only H$\alpha$ star-forming galaxies (in the cluster, outskirts or field), we find cluster star-forming galaxies to be significantly metal rich, with a median metallicity 12\,+\,log(O/H)\,=$8.611\pm0.004$, which compares with 12\,+\,log(O/H)\,=$8.476\pm0.003$ for the cluster outskirts and 12\,+\,log(O/H)\,=$8.433\pm0.016$ for outside the cluster. Our results thus clearly indicate that star-forming galaxies in the merging clusters are significantly metal rich, practically solar, being about $\sim0.15$\,dex more metal rich than other star-forming galaxies outside the cluster.

We find that star-forming galaxies in the cluster show significantly higher metallicities than star-forming galaxies in the outskirts or in the field environment, although star-forming galaxies in the post-shock regions show an even higher metallicity, fully consistent with solar metallicity. We note, however, that star-forming galaxies in the post-shock regions also have a slightly higher median stellar mass (+0.12\,dex), and thus the slightly higher metallicities when compared with the remaining galaxies in the cluster, can be fully explained by the mass-metallicity relation (see \S\ref{Mass_metal}). Thus, both sub-samples show a very high metal-enrichment.

We note that star-forming galaxies in the cluster are also slightly more massive, as a whole, than those in the other environments, and thus it is very important to look at the mass-metallicity relation, in order to address whether the higher metallicity is simply a consequence of higher stellar masses, or a genuine higher metallicity even at fixed mass.

\subsection{The Mass-Metallicity relation for the Sausage merging cluster} \label{Mass_metal}

Having found that our merging cluster star-forming galaxies have higher metallicities than those in lower density environments, we investigate the mass-metallicity relation. We show our results in Figure \ref{Mass_metallicity}. We find a strong relation between metallicity (here traced by the [N{\sc ii}]\,/\,H$\alpha$ ratio and using the conversion of \citealt{Pettini04} for star-forming galaxies) and stellar mass, both when we look at individual sources (Figure \ref{Mass_metallicity}: right panel) and particularly when we look at stacks as a function of stellar mass (Figure \ref{Mass_metallicity}: left panel). We find that H$\alpha$ emitters in both the cluster and the field have metallicities that correlate with stellar mass. We also show where AGNs would be placed had they not been excluded from our analysis, clearly showing that they would bias the metallicities to higher values. We note that all AGNs were excluded from the metallicity analysis, both for the fits with individual sources and for all the stacks that measured metallicities (the only exceptions are for ``full" samples in Table 2, but we make explicit notes that those values are still contaminated by AGN). We also note that because of significant dust extinction on the line of sight, the [O{\sc iii}]\,5007/H$\beta$ ratio is overestimated, thus making our cuts even more conservative in excluding potential AGN.

Our results clearly reveal, both based on the combination of individual measurements, and based on the stacks for each sub-sample, that our merging cluster star-forming galaxies are significantly more metal-rich than those outside the cluster. We find this to be valid for masses higher than 10$^9$\,M$_{\odot}$ (for which we are reasonably complete), although the difference seems to be even higher for masses $>10^{10}$\,M$_{\odot}$. We note that this difference, of about 0.15 to 0.2\,dex, found at all masses, is based on two fully comparable samples, with the same selection function, same completeness, with the sole difference being the environment in which these star-forming galaxies reside in. We also compute metallicities using the O3N2 indicator \citep{Alloin79} and recover similar results. However, measurements based on O3N2 have much lower S/N, due to the high level of dust extinction affecting both [O{\sc iii}]\,5007 and H$\beta$ (up to $\sim3$\,mag). In addition, because of the difference in wavelength between the two lines ($\sim176$\,\AA), the [O{\sc iii}]\,5007 and H$\beta$ emission lines suffer from different dust extinction values. Therefore, the [O{\sc iii}]\,5007/H$\beta$ line ratio is biased high (average $\sim0.06$\,dex), while [N{\sc ii}]\,/\,H$\alpha$ ratio provides, in this case, a much better metallicity estimator. With [N{\sc ii}]\,/\,H$\alpha$ we can measure metallicities at much higher S/N, in exactly the same way for our fully comparable samples in and outside the cluster and without the potential biases from dust extinction, as [N{\sc ii}] and H$\alpha$ are only separated by 20.8\,\AA.

We also compare our results with those in the literature, particularly with SDSS \citep[][after applying the appropriate corrections for a different metallicity indicator and a different IMF]{Maiolino08}. In practice, we find that star-forming galaxies in the Sausage merging cluster follow the local ($z\sim0$) mass-metallicity perfectly, even though they are being studied 2.3\,Gyr before it was established. On the other hand, star-forming galaxies outside the cluster follow a mass-metallicity relation more applicable to slightly higher redshift galaxies \citep[][]{Maiolino08}. We also use the parameterisation of \cite{Maiolino08} to fit our mass-metallicity relations for cluster and star-forming galaxies outside the cluster. The parameterisation is given by: 12\,+\,log(O/H)\,=$-0.0864\times(\log\,M-M_0)^2+K_0$. For Cluster star-forming galaxies we find $M_0=10.68\pm0.04$ and $K_0=8.72\pm0.01$, while for star-forming galaxies outside the cluster the best fit is given by $M_0=10.49\pm0.14$ and $K_0=8.56\pm0.03$.

While we find evidence that the Fundamental Metallicity Relation \citep{Mannucci10} is somewhat applicable to our data (at all environments), our sample (particularly when split in different environments and only focusing on robust star-forming galaxies) is too small to properly address how these sources fit into the FMR and particularly to attempt to constrain it. However, we note that both samples (cluster and outside the cluster) are very well matched in SFR (see Figure \ref{SFR_vs_MASS} and \cite{Stroe15}), and thus the difference in metallicity for a fixed mass cannot be explained by a typically lower SFR. Nevertheless, it should be noted that the scatter on individual measurements in Figure \ref{Mass_metallicity} (right panel) seems to be mostly driven, at fixed stellar mass and fixed environment, by SFR.

\section{Discussion: Shock induced star-formation, cooling or turbulence?}

By obtaining high S/N spectra of the bulk of the sample of candidate line emitters in the Sausage cluster, we were able to confirm them as H$\alpha$ emitters. We find that about 65\% are consistent with being powered by star-formation, with about 35\% being AGN. We find H$\alpha$ star-forming galaxies in the cluster to be highly metal-rich and to already follow the SDSS $z\sim0$ mass metallicity relation. We also find striking evidence of ubiquitous outflows in the majority of our cluster H$\alpha$ emitters: not only strong P Cygni profiles, mostly in cluster AGN, but also for star-forming galaxies, where we find redshifted emission lines and particularly significantly blue-shifted Na\,D emission. We find that such outflows are consistent with being driven by AGN for sources with clear AGN activity, while for star-forming galaxies in the cluster, and particularly for those with very high [S{\sc ii}]$_{6716}$/H$\alpha$, away from the post-shock regions, these are likely driven by supernova. It is also likely that star-forming galaxies in the post-shock region are in a relatively earlier evolution phase compared to those away from it (which are likely in final phase of star-formation, showing the strongest outflows and the strongest evidence for supernova). We argue that the merger must have had a significant effect on all these H$\alpha$ emitters. This is because the cluster, despite being extremely massive, shows a surprising number of active H$\alpha$ emitters, but also because all H$\alpha$ emitters in the cluster show significant differences in their properties to field galaxies.

A requirement for the shock and cluster merger to increase star formation and AGN activity is that the galaxies within the sub-clusters are still relatively gas rich or have at least some remaining amount of relatively cool molecular gas, capable of being turned into stars in a few Myrs, or be accreting such gas at a sufficient rate. The H$\alpha$ emitters in the `Sausage' cluster present masses $10^{9-10.7}$\,M$_\odot$ and are in general very metal-rich, particularly given their mass, following the SDSS relation at $z=0$. Field H$\alpha$ emitters at the same redshift but outside the cluster show systematically lower metallicities at all masses (see Figure~\ref{Mass_metallicity}). The metallicity as measured from nebular lines for the H{\sc ii} regions is essentially solar for cluster star-forming galaxies, suggesting that these H$\alpha$ emitters are using relatively metal rich gas to form new stars at all stellar masses. What is the source of these reservoirs of gas?

A source of gas would be a reservoir of high-metallicity $Z\sim0.3$ ICM gas \citep{Leccardi08}, compared to field galaxies which may preferentially accrete low(er)-metallicity gas \citep[$Z\sim0.01$ at $z\sim0.2$,][]{Fox11} from their inter-galactic (filamentary) medium. Accretion of ICM gas was also proposed as an interpretation for the metal-rich ($Z\sim1.1Z_\odot$) spirals found in the Virgo cluster \citep{Skillman96}. By contrast, if we assume the galaxies to be closed-boxes, supernova explosions (SN) of asymptotic giant branch (AGB) stars would enrich the intra-galactic medium with metals and, given the higher-mass of the galaxies and the large potential of the cluster, this gas could be retained and fall back into the galaxies.

A slight elevation of $0.04$ dex was also found in the metallicities of a large sample of cluster galaxies, as compared to the field \citep[][]{Ellison09} -- our results go in the same direction, but we find an even higher offset. \cite{Cooper08} also find that galaxies at low redshift residing in higher density environments tend to have higher metallicities, at fixed mass, than those in lower density regions, in agreement with our findings. Interestingly, this trend is also being found at higher redshift. By studying an over-density of H$\alpha$ emitters at $z=0.8$ with KMOS, \cite{Sobral13B} find that star-forming galaxies in the high-density group-like or filamentary structure are more metal rich than those in the field. However, the difference can be explained by the fact that galaxies residing in higher density regions are also more massive. On the other hand, and at higher redshift, \cite{Kulas13} used MOSFIRE to study a ``proto-cluster'' at $z\sim2.3$. They also find that galaxies in the proto-cluster environment are, on average, more metal rich than those in the field comparison (which the authors also obtain with the same instrument and set-up, to be fully comparable), particularly for stellar masses of $\sim10^{10}$\,M$_{\odot}$. Similar results are found by \cite{Shimakawa15}, who study two rich over-densities at $z\sim2.2$ and $z\sim2.5$. \cite{Shimakawa15} find that galaxies residing in over-densities likely have higher metallicities than those in the field sample presented by \cite{Erb2006}.

As long as there is some relatively cool gas in cluster galaxies, and even if that gas is relatively unlikely to form stars on its own (e.g. not dense enough/too stable), the passage of a shock wave can likely introduce the turbulence needed for that to happen. Given the shock properties and velocity, the shock is expected to traverse galaxies within a relatively very short timescale of about $10-50$ Myr. Hence the shock induces turbulence quickly, which may lead to further gas cooling and collapse of any gas that is still available in the galaxies -- although due to the time needed for that to happen, a time delay is expected from the passage of the shock wave to the star-formation episodes. However, the enhancement of star formation and AGN activity following the shock passage can quickly deplete the gas reservoir. This is because while part of the gas fuels SF and goes into stars, we also find evidence of strong outflows in our cluster H$\alpha$ star-forming galaxies, and also for our H$\alpha$ cluster AGN: these can easily further remove gas and lead to relatively short depletion times. We therefore expect the passage of the shock to lead to a steep rise in SF for a few $10-100$ Myr, followed by a quick quenching of the galaxy and a shut-down in the formation of new stars. Given the evidence for strong outflows and supernova in cluster H$\alpha$ star-forming galaxies not in the post-shock region (which may have been affected even longer ago), such galaxies may be in the final phase of quenching. This is a very likely scenario, particularly because the latter are satellites of extremely massive dark matter haloes of $>10^{15}$\,M$_{\odot}$. Therefore, any gas that is expelled from the galaxy by strong outflows will easily be lost to the ICM.

We also note that a high number of our H$\alpha$ emitters in the Sausage cluster are located near the shock fronts, in the post-shock region, fully consistent with the shock front affecting them 100-200\,Myrs ago. At the passage of the shock wave two potentially important things happen: i) firstly, magnetic fields are amplified and aligned and they funnel material to infall only along the field lines (this may have helped sources to accrete ICM gas in some conditions and/or to force gas in the galaxies to become denser) and ii) after the shock passes, turbulence takes over and the fields also get tangled; thus, such conditions (provided galaxies still have some molecular gas) should enhance/promote star formation. 

We therefore conclude that whatever process is driving the enhanced star-formation activity in the merging cluster, it will contribute to the build-up of the red sequence, as even though new stars will form, the feedback processes that we see happening will quickly quench any galaxy that still had enough gas to form stars and that was able to cool/accrete gas.

\section{Conclusions}

We presented spectroscopic observations of 83 strong H$\alpha$ emitters in the ``Sausage" merging cluster and in surrounding regions. Our sample, split into cluster, outskirt and field H$\alpha$ emitters, selected in the same way, and with very high S/N, allows us to unveil the nature and properties of sources, and directly compare them across environment. Our main results are:

\begin{itemize}

\item  We find that $\sim35$\,\% of the cluster H$\alpha$ emitters are AGN, similar to what is found in the field ($29\pm7$\%). We do not find any significant evidence for galaxy-galaxy (major) mergers in our H$\alpha$ emitters in the cluster, thus ruling out that the elevated activity is due to galaxy-galaxy mergers.

\item Cluster star-forming galaxies in the hot X-ray gas and/or in the cluster sub-cores show exceptionally high [S{\sc ii}]\,6716, implying very low electron densities ($<50\times$ lower than all other star-forming galaxies) and/or significant contribution from supernova.

\item Cluster star-forming galaxies show evidence of significant outflows (blueshifted NaD, $200-600$\,km\,s$^{-1}$), likely driven by supernova. Individual signatures of strong, massive outflows are also found for the cluster H$\alpha$ AGN, including P Cygni profiles. All cluster star-forming galaxies near the centre of the merging cluster show significant outflows, and thus this will likely lead to star-formation being quenched rapidly.

\item Cluster star-forming galaxies are highly metal-rich, roughly solar, and those in the post-shock region are the most metal rich (12 + log(O/H)$=8.632\pm0.004$).

\item H$\alpha$ star-forming galaxies in the Sausage merging cluster follow the local Universe mass-metallicity relation. However, H$\alpha$ star-forming galaxies in the Sausage merging cluster also show systematically higher metallicity ($\sim$0.15-0.2\,dex) for $M>10^{9}$\,M$_{\odot}$ when directly comparing with our H$\alpha$ emitters outside the cluster. This suggests that the shock front may have triggered remaining gas which galaxies were able to retain into forming stars.

\end{itemize}

Our observations show that the merger of massive ($\sim10^{15}$\,M$_\odot$) clusters can provide the conditions for significant star-formation and AGN activity, but, as we witness strong feedback by star-forming galaxies and AGN (and given how massive the merging cluster is), and particularly because these sources reside in very massive haloes of $>10^{15}$\,M$_{\odot}$ which will not likely allow galaxies to re-accrete gas, such sources will likely be quenched in a few 100\,Myrs.

\section*{acknowledgments}

We thank the referee for many helpful comments and suggestions which greatly improved the clarity and quality of this work. DS acknowledges financial support from the Netherlands Organisation for Scientific research (NWO) through a Veni fellowship, from FCT through a FCT Investigator Starting Grant and Start-up Grant (IF/01154/2012/CP0189/CT0010) and from FCT grant PEst-OE/FIS/UI2751/2014. AS and HR acknowledge financial support from an NWO top subsidy (614.001.006). R.J.W. is supported by NASA through the Einstein Postdoctoral grant number PF2-130104 awarded by the Chandra X-Ray Center, which is operated by the Smithsonian Astrophysical Observatory for NASA under contract NAS8-03060. Part of this work performed under the auspices of the U.S. DOE by LLNL under Contract DE-AC52-07NA27344. Some of the data presented herein were obtained at the W.M. Keck Observatory, which is operated as a scientific partnership among the California Institute of Technology, the University of California and the National Aeronautics and Space Administration. The Observatory was made possible by the generous financial support of the W.M. Keck Foundation. This research has made use of the NASA/IPAC Extragalactic Database (NED) which is operated by the Jet Propulsion Laboratory, California Institute of Technology, under contract with the National Aeronautics and Space Administration. This research has made use of NASA's Astrophysics Data System. Dedicated to the memory of C.\,M. Sobral (1953-2014).

\bibliographystyle{mn2e.bst}
\bibliography{bibliography.bib}

\bsp

\label{lastpage}

\end{document}